%% file: main.tex
\documentclass[10pt,journal,compsoc]{IEEEtran}
\usepackage{microtype}                 
\PassOptionsToPackage{warn}{textcomp}  
\usepackage{textcomp}                  
\usepackage{mathptmx}                  
\usepackage{times}                     
\usepackage{tabu}                      
\usepackage{booktabs}                  
\usepackage{cite}

\usepackage{paralist}
\usepackage{amssymb}
\usepackage{float}
\usepackage{placeins}
\usepackage{scrextend}
\usepackage{graphicx}
\usepackage[table,xcdraw]{xcolor}
\usepackage{makecell}

\usepackage[draft=true]{minted}
\usepackage[hyphens]{url}
\usepackage[hidelinks]{hyperref}
\hypersetup{breaklinks=true}
\usepackage{array}
\newcolumntype{P}[1]{>{\centering\arraybackslash}p{#1}}

\usepackage{ulem}

\usepackage{xspace,xpunctuate}
\newcommand{\ie}{{i.e.,}\xspace}
\newcommand{\eg}{{e.g.,}\xspace}
\newcommand{\ea}{{et~al\xperiod}\xspace}

\newcommand{\npaper}{98}
\newcommand{\aivis}{\textsc{AI4VIS}}

\newcommand{\webLink}{\href{https://ai4vis.github.io}{ai4vis.github.io}}

\usepackage{tikz}
\definecolor{nodeBG}{RGB}{29,40,97}
\newcommand*\circled[1]{\tikz[baseline=(char.base)]{
            \node[shape=circle,draw, fill=nodeBG,inner sep=0.5pt] (char) {\textcolor{white}{#1}};}}
            
\definecolor{tableBG}{RGB}{221,221,221}

\newenvironment{revised}[0]{%
    \leavevmode\color{black}\ignorespaces
}{}

\ifCLASSINFOpdf
\else
\fi
\begin{document}
%
\title{AI4VIS: Survey on Artificial Intelligence Approaches for Data Visualization}
%
%
%
%

\author{Aoyu~Wu,~\IEEEmembership{}
        Yun~Wang,~\IEEEmembership{}
        Xinhuan~Shu,~\IEEEmembership{}
        Dominik~Moritz,~\IEEEmembership{}
        Weiwei~Cui,~\IEEEmembership{}
        Haidong~Zhang,~\IEEEmembership{} \\
        Dongmei~Zhang,~\IEEEmembership{}
        and~Huamin~Qu~\IEEEmembership{}
\IEEEcompsocitemizethanks{\IEEEcompsocthanksitem A. Wu, X. Shu, and H. Qu are with the Hong Kong University of Science and Technology. Email: \{awuac, xinhuan.shu, huamin\}@cse.ust.hk. This work is done when A. Wu is an intern at MSRA.
\IEEEcompsocthanksitem Y. Wang, W. Cui, H. Zhang, and D. Zhang are with Microsoft Research Asia. Email: \{wangyun, weiwei.cui, haidong.zhang, dongmeiz\}@microsoft.com.
\IEEEcompsocthanksitem D. Moritz is with Carnegie Mellon University and Apple. Email: domoritz@cmu.edu.}
\thanks{Manuscript received xx xxx. xxxx; revised xx xxx. xxxx; accepted xx xxx. xxxx. Date of publication xx xxx. xxxx; date of current version x xxx. xxxx. (Corresponding author: Yun Wang).}
\thanks{Recommended for minor revision}
\thanks{Digital Object Identifier:}
}

%
%

\markboth{IEEE Transactions on Visualization and Computer Graphics,~Vol.~xx, No.~xx, xx~2018}%
{}
%



\input{section/sec0Abstract}

\maketitle

\IEEEdisplaynontitleabstractindextext

%
\IEEEpeerreviewmaketitle

\input{section/sec1Intro}
\input{section/sec2RelatedWork}
\input{section/sec3Back}

\input{section/sec5Data}

\input{section/sec4Why}
\input{section/sec6Task}

\input{section/sec7FutureWork}

\input{section/sec8Discussion}

\input{section/sec9Conclusion}


%



\ifCLASSOPTIONcompsoc
  \section*{Acknowledgments}
\else
  \section*{Acknowledgment}
\fi

The authors would like to thank anonymous reviewers for their constructive comments.
This research is supported by Hong Kong Theme-based Research Scheme Grant T41-709/17N.

\ifCLASSOPTIONcaptionsoff
  \newpage
\fi



\bibliographystyle{IEEEtran}
\bibliography{main}



%

\input{section/biography}
\end{document}

%% file: section/sec0Abstract.tex
\IEEEtitleabstractindextext{%
\begin{abstract}
Visualizations themselves have become a data format.
Akin to other data formats such as text and images,
visualizations are increasingly created, stored, shared, and (re-)used with artificial intelligence (AI) techniques.
In this survey,
we probe the underlying vision of formalizing visualizations as an emerging data format
and review the recent advance in applying AI techniques to visualization data (AI4VIS).
We define visualization data as the digital representations of visualizations in computers and focus on data visualization (\eg charts and infographics).
We build our survey upon a corpus spanning ten different fields in computer science with an eye toward identifying important common interests.
Our resulting taxonomy is organized around 
WHAT is visualization data and its representation,
WHY and HOW to apply AI to visualization data.
We highlight a set of common tasks that researchers apply to the visualization data and present a detailed discussion of AI approaches developed to accomplish those tasks.
Drawing upon our literature review,
we discuss several important research questions surrounding the management and exploitation of visualization data,
as well as the role of AI in support of those processes.
We make the list of surveyed papers and related material available online at \webLink.



\end{abstract}

\begin{IEEEkeywords}
Survey; Data Visualization; Artificial Intelligence; Data Format; Machine Learning
\end{IEEEkeywords}}

%% file: section/sec1Intro.tex
\IEEEraisesectionheading{\section{Introduction}\label{sec:introduction}}
\IEEEPARstart{D}{ata} visualizations use visual representations of abstract data to amplify human cognition.
Researchers traditionally investigate visualizations as artifacts created for people.
This paper revisits this traditional perspective in line with the growing research interest in applying artificial intelligence~(AI) to visualizations.
Similar to common data formats like text and images,
visualizations are increasingly created, shared, collected, and reused with the power of AI.
Thus,
we see that visualizations are becoming a new data format processed by~AI.
For instance, this trend was evident at the 2020 IEEE Visualization Conference, where multiple techniques were proposed for automating the creation of visualizations~\cite{qian2020retrieve,wu2020mobilevisfixer,oppermann2020vizcommender,shi2020calliope,kim2020gemini},
retargeting visualizations~\cite{zhang2020viscode,fu2020chartem}, and analyzing visualization ensembles~\cite{chen2020composition,zhao2020chartseer}.
In light of this trend, 
new concepts and research problems are emerging,
raising the need to organize existing literature and clarify the research landscape.

\begin{figure*}[htp]
    \setlength{\abovecaptionskip}{0pt}
	\centering
	\includegraphics[width=1\linewidth]{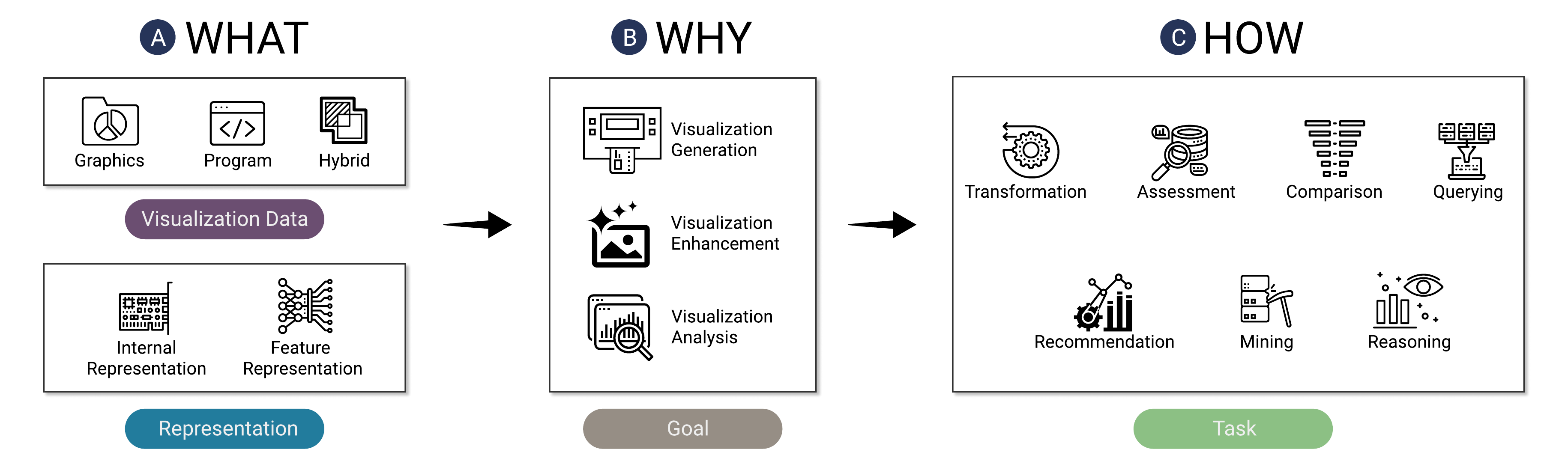}
	\caption{We categorize the surveyed papers from three aspects: {\protect\circled{A}} What is the visualization data and its representation; {\protect\circled{B}} Goals for why apply AI to visualization data; and {\protect\circled{C}} How to apply AI to visualization data, where we outline seven tasks and separately discuss corresponding AI approaches.}
	\label{fig:overview}
	\vspace{-1em}
\end{figure*}

This survey describes the research vision of 
formalizing visualizations as an emerging data format
and reviews recent advances in developing AI approaches for visualization data~(\aivis{}).
We define \textbf{visualization data} as the digital representations of visualizations in computers and focus on data visualization.
Different from common formats,
visualization data contains multimodal information such as visual encodings, encoded data, text, and images.
Those characteristics pose new challenges in developing tailored AI approaches (\eg how to represent visualization data).

Nevertheless,
AI is a broad notion that has been studied in different areas.
Those areas have different motivations and research questions about applying AI to visualization data,
the proposed techniques,
and the content format used to represent visualizations.
For instance,
the \textit{Web and Information Retrieval} community advances the technology for searching visualizations (\eg~\cite{chen2015diagramflyer,ray2015architecture}),
while research in \textit{Computer Vision} recently studies visual question answering in charts (\eg~\cite{kafle2018dvqa,chen2020figure}).
Therefore,
a comprehensive understanding of~\aivis{} research requires a foundation in rich literature from diverse disciplines.

We construct a literature corpus by a relation-search approach~\cite{mcnabb2017survey}, 
\ie graph traversal over the citation and reference networks.
This approach allows us to collect~\npaper{} publications from 10 research communities -- publications from the visualization community account for roughly one-third.
While not being exhaustive,
our corpus provides sufficient research instances to synthesize relevant work that contributes to the understandings of important common research questions and techniques.
As shown in~\autoref{fig:overview},
we organize and categorize \aivis{} research following a well-established what-why-how viewpoint~\cite{xu2020survey}:

\begin{compactitem}
    \item \textbf{What} is visualization data. We formalize the concept of visualization data by providing an overview of its content formats as well as the representations~(\autoref{fig:overview}\circled{A}). 
    \item \textbf{Why} apply AI to visualization data. We identify three common goals for~\aivis{} research, 
    namely visualization generation, enhancement, and analysis (\autoref{fig:overview}\circled{B}).
    We classify those goals into subcategories to provide a comprehensive response to ongoing discussions like ``why should we teach machines to read charts made for humans''~\cite{ono2018should}.
    \item \textbf{How} to apply AI to visualization data. Most importantly,
    we contribute a task abstraction regarding how to apply AI to visualization data (\autoref{fig:overview}\circled{C}).
    The task abstraction is critical since it allows for domain-agnostic and consistent descriptions of research questions across different disciplines.
    Besides,
    it facilitates an organized discussion reconciling distinct visualization paper types~\cite{lee2019broadening},
    \ie a \textit{technique} paper focuses on a single task, while a \textit{system} paper might accomplish multiple tasks.
    In total, we note seven common tasks and discuss the AI approaches for each task separately.
\end{compactitem}

Drawing upon the discussion, we outline research opportunities.
We make the list of surveyed papers with related material available online at \href{https://ai4vis.github.io}{\mintinline{html}{ai4vis.github.io}}.
We hope that our survey will
stimulate new theories, problems, techniques, and applications in this growing research area.

%% file: section/sec2RelatedWork.tex
\section{Related Surveys}
Our ability to collect data has significantly exceeded our ability to analyze it,
contributing to the emergence of AI approaches that automate the processes.
Recently,
there are active ongoing discussions about how visualization research could be interwoven with artificial intelligence (AI)~\cite{visai,andrienko2020big}.
However, 
AI has been defined and operationalized as a broad notion.
To clarify our discussion,
our specific prospect is to formalize visualizations as an emerging format of data.
We see visualizations different from existing types such as images and texts,
thereby raising many research questions regarding how AI facilitates the manipulation and analysis of visualizations.

Several surveys review techniques for automating the creation of visualizations.
Saket \ea~\cite{saket2018beyond} discussed the prospect of learning visualization design and classified automated visualization design systems into knowledge-based (\ie rule-based), data-driven (\ie machine-learning), and hybrid approaches.
This classification was systematically reviewed in a recent survey about visualization and infographics recommendation~\cite{zhu2020survey}.
Besides, Qin~\ea~\cite{qin2020making} drew on research in the database community to survey what makes data visualization more efficient and effective.
In addition to automated creation and recommendation,
Davila~\ea~\cite{davila2020chart} reviewed research over the past eight years to formalize \textit{chart mining}, defined as ``the process of automatic detection, extraction and analysis of charts to reproduce the tabular data that was originally used to create them''.
Wang~\ea~\cite{wang2020applying} recently surveyed machine learning models applied to visualizations.
Different from them,
our objective is to provide a comprehensive review of~\aivis{} research, 
that is, a scaffold for emerging research problems to be formulated and understood (\eg automatic assessment and summarization of visualizations).

%% file: section/sec3Back.tex
\section{Methodology}
In this section, we describe the scope of our literature survey, the search methodology and corpus, as well as our analysis method.

\subsection{Definition and Scope}
This survey focuses on AI approaches applied to visualization data.
We define visualization data as the digital representations of visualizations in computers.
Thus,
we include existing work that contributes AI techniques or systems that primarily focuses on \textit{inputting or outputting visualization data}.
However,
due to the wide scope of visualizations and related research,
we constrict the scope of visualizations for manageability.

\textbf{Excluding scientific visualizations.}
We cover literature related to information visualization and visual analytics, 
particularly for which visualizations are represented as charts and infographics.
This restriction excludes literature that primarily focuses on scientific visualizations.
Scientific visualizations represent scientific data such as flows and volumes, which are typically designed with a strong inherent reference to space and time~\cite{kehrer2012visualization}.
We exclude them due to their heterogeneous nature that yields different research challenges and interests~\cite{xu2020survey}.

\textbf{Excluding research specific to a chart type.}
The research problems and proposed techniques could vary among different chart types.
For instance,
a conference (\ie International Symposium on Graph Drawing) in the graph visualization community focuses on the graph layout problem,
whereas their problems might not apply to other visualizations like histograms.
Those chart-specific studies are large-scale and therefore might not be covered in a single survey,
\eg
Behrisch~\ea~\cite{behrisch2018quality} devoted a survey to discussions about assessment metrics for different charts.
Different from them,
our goal is to identify common research problems (\ie assessment, recommending) irrespective of the chart types.
Nevertheless,
we note a few chart-type-specific studies and discuss how they might fall into our taxonomy in~\autoref{sec:discussion}.

\textbf{Excluding human interaction data.}
Finally,
we emphasize the central role of visualization data in the surveyed AI approaches.
In other words,
we do not consider work whose primary goal is to collect and analyze human interaction data when creating or using visualizations.
Therefore,
we exclude provenance data~\cite{xu2020survey} and data collected for natural language interfaces~\cite{srinivasan2017natural}.


\subsection{Method and Corpus}
To establish the corpus of papers we discussed in this survey,
we apply a relation-search method~\cite{mcnabb2017survey} to traverse the literature.
Our method starts with a linear scan of full papers published at the 2020 IEEE Visualization Conference to collect the starting points. 
This initial set of papers has nine papers~\cite{qian2020retrieve,wu2020mobilevisfixer,oppermann2020vizcommender,shi2020calliope,kim2020gemini,zhang2020viscode,fu2020chartem,chen2020composition,zhao2020chartseer}.
We further augment these starting points with papers covered in related surveys~\cite{saket2018beyond,davila2020chart,zhu2020survey,qin2020making}.
If a paper is selected to be included,
we traverse both its references and citations.
We search for papers breadth-first in an attempt to avoid over-focus on one particular line of research. 

\begin{figure}[!t]
    \setlength{\abovecaptionskip}{0pt}
	\centering
	\includegraphics[width=1\linewidth]{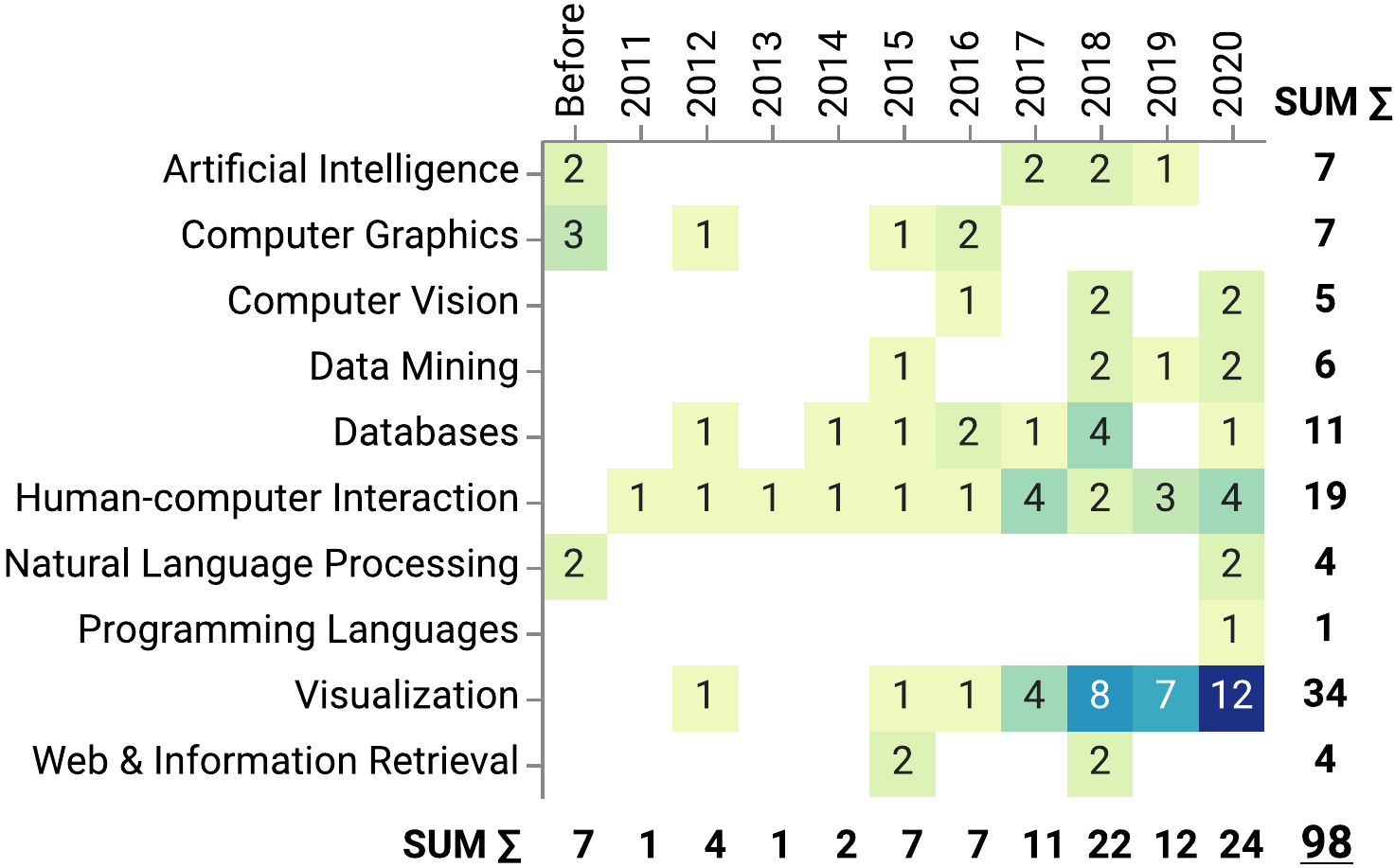}
	\caption{The number of papers per research area and per year.}
	\label{fig:paperStat}
	\vspace{-1em}
\end{figure}

Our relation-search method results in an interdisciplinary corpus consisting of \npaper{} papers from 10 research areas.
We decide the research area of each paper according to the classification of computer science areas by CSRankings\footnote{http://csrankings.org/}.
\autoref{fig:paperStat} lists the research areas,
indicating the interdisciplinary nature and widespread popularity of visualizations.
The primary area is Visualization,
accounting for around one-third (34/98).
The next areas are Human-computer Interaction and Databases.  
Besides, we see research efforts from Artificial Intelligence, Data Mining, Computer Graphics, and the Web \& Information Retrieval. 

Another finding is the increasing trend of applying AI methods for data visualization. 
\autoref{fig:paperStat} shows that the total number of publications has been increasing steadily over the last decade,
particularly with a surge since 2018 and a peak at 2020.
Given the wide and ever-growing research efforts and interests,
we believe that this topic will receive more attention in the near future.

We acknowledge the limitations of our search methodology which is based on manual search over citation and reference graphs.
Consequently,
our survey should be seen as an effort in investigating the diversity of related research,
providing sufficient research instances to contextualize the current research landscape,
and indicating future research opportunities. 
As such,
we do not claim comprehensiveness nor exhaustiveness.
Instead,
future research could augment the corpus by automated graph traversal.

\subsection{Coding and Classification}
\autoref{fig:overview} provides an overview of our classifications of surveyed papers,
which is organized according to the what-why-how axes.
This organization is well-established in visualization-related surveys~\cite{xu2020survey,behrisch2018quality}.
Nevertheless,
we introduce two modifications to the ``what'' and ``how'' axes in order to contextualize our discussion in~\aivis{} research.

Firstly, from the ``what'' perspective,
we discuss not only what visualization data is but also its representation (\autoref{sec:what}).
For representation,
we review internal representations (how visualization data is stored and operated in systems) and feature representation (how visualizations are converted into features that are mathematically and computationally convenient for analysis).

Secondly, from the ``how'' perspective,
we organize approaches with a novel task abstraction.
We identify seven common tasks for~\aivis{} research and discuss approaches for each task in \autoref{sec:how}.
Our motivations for such task abstractions are two-fold:
\begin{compactitem}
    \item \textbf{Reconciling techniques with system papers}. We find our corpus a mixture of technique and system papers that confound the discussion. A technique paper typically solves a single task, while a system paper could consist of multiple components, each for different tasks~\cite{lee2019broadening}. As such, we aim to decompose system papers into abstract tasks that allow for a collectively exhaustive taxonomy of tasks.
    \item \textbf{Unifying Inconsistent Vocabularies}. Due to the interdisciplinary nature of our corpus, we find tasks are usually described in inconsistent vocabularies. For instance, the task of extracting encoding choices from visualization images or specifications is described as deconstruction~\cite{wu2020mobilevisfixer,qian2020retrieve} or chart mining~\cite{davila2020chart}. Thus, we wish to establish a common vocabulary that enables consistent discussions for researchers from different areas to communicate the relevance and subtleties.
\end{compactitem}

For the above purposes,
we adopted a bottom-up approach by iteratively categorizing and labeling the tasks in surveyed papers.
Thanks to the interdisciplinary nature of our corpus,
we were able to start from several ``seeding'' tasks by referring to task taxonomy in other fields~\cite{computerVision}.
Subsequently,
two authors labelled and verified whether the task could map to existing categories,
and if not applicable,
discussed alternative task categorizations with the remaining authors.
Conflicts of labels were also discussed until reaching a consensus.
During the labeling process,
we found that most tasks conformed to the ``seeding'' terms with few exceptions that reflected the peculiarity of visualizations.
For exceptions,
we discussed candidate terms in surveyed literature to select tasks that were understandable and widely communicated (\eg recommendation).
We discuss details of the task taxonomy in \autoref{sec:how}.

Our supplemental website, available at \webLink, provides the details of our labels.
Particularly,
we label tasks at the (sub-)section level,
and provide quotes from the paper to help readers understand why it falls into the task category.
The website also provides an overview of the surveyed paper.
\begin{revised}
In the following text,
citations refer to all instances for the reported category,
except where noted otherwise by phrases such as ``for instance'' or ``like''.
\end{revised}

%% file: section/sec5Data.tex
\section{Data: What is Visualization Data}
\label{sec:what}
\begin{figure}[!t]
    \setlength{\abovecaptionskip}{5pt}
	\centering
	\includegraphics[width=1\linewidth]{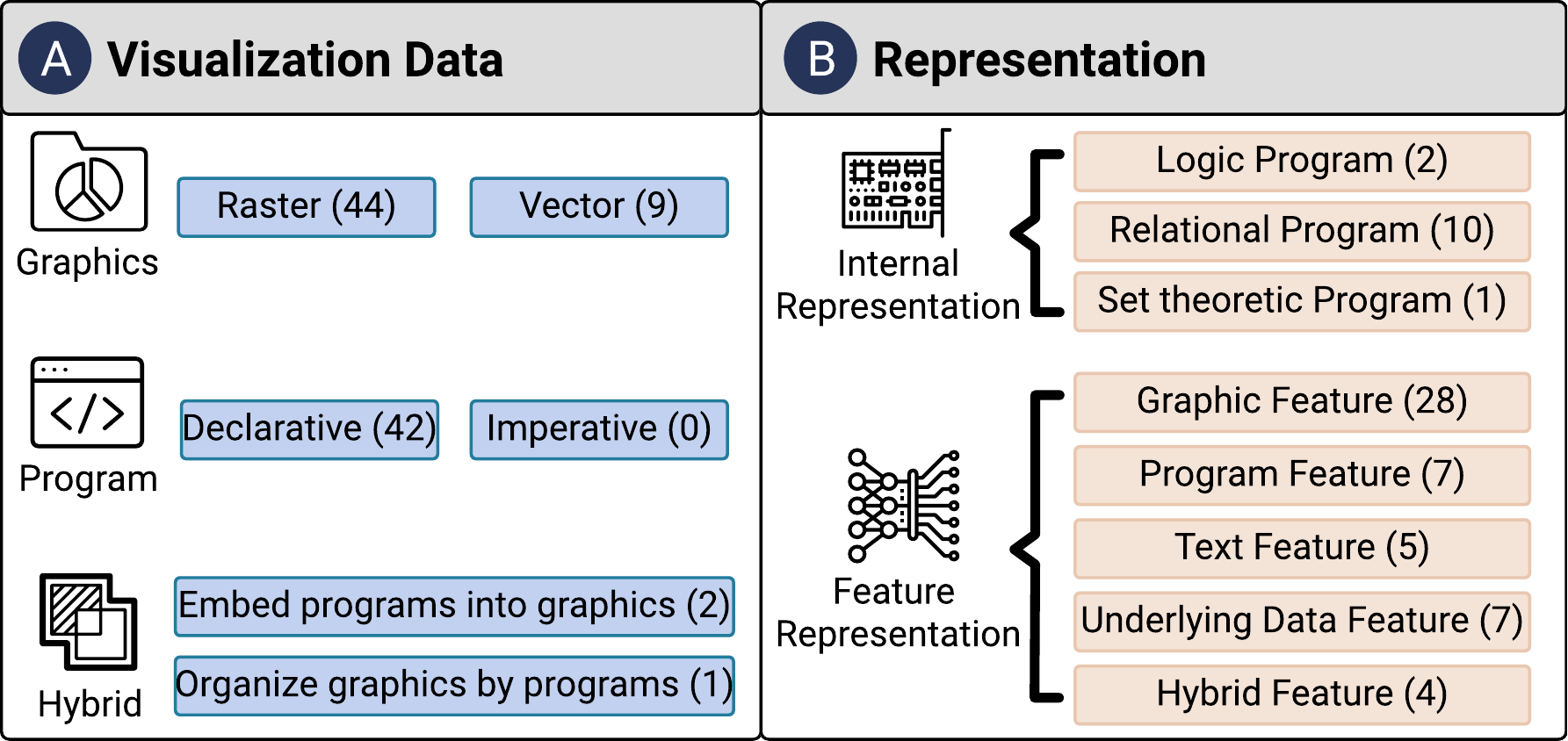}
	\caption{Summary of categorization of~{\protect\circled{A}} visualization data and~{\protect\circled{B} its representation}. Numbers in the parentheses indicate the count.}
	\label{fig:data}
	 \vspace{-1em}
\end{figure}

In this section,
we formalize the concept of visualization data.
Specifically,
we discuss and categorize visualization data in terms of its raw data format (\autoref{sec:data:rawdata}) and representations (\autoref{sec:data:representation}).
As shown in~\autoref{fig:data},
we classify visualization data formats into \textbf{graphics}, \textbf{programs},
and \textbf{hybrid} that blends the benefits of both.
In addition to raw data,
we note visualizations are sometimes represented as carefully designed \textbf{internal representation} formats in surveyed systems.
Internal representations are usually proposed to facilitate the computing by removing unnecessary information,
\eg
the VQL format~\cite{luo2020interactive} only stores data transformation and encoding without style information.
As such,
internal representations are usually not exposed (outputted and shared).
Finally,
we review feature presentations, including \textbf{feature engineering} and \textbf{feature learning}.
Feature presentations are vital for machine learning tasks by concerting visualizations into features that are mathematically and computationally convenient to analyze.
We discuss them due to the increasing interest in applying machine learning to visualizations.

\input{section/data/D1RawData}
\input{section/data/D2Feature}

%% file: section/data/D1RawData.tex
\subsection{Data}
\label{sec:data:rawdata}
Visualization data can be stored in different content formats such as graphics and programs.
The choice of content formats directly influences the downstream operation possibly on visualization data,
since different formats have own advantages and disadvantages.
Here,
we discuss three formats we identified in our corpus: graphics, programs, and hybrids.
As shown in~\autoref{fig:data},
both graphics and programs are common in the corpus,
and three recent research papers has proposed hybrid formats for the combined power.

\subsubsection{Graphics}
\textbf{Graphics} are a natural and expressive content format of visualizations, 
since visualizations are defined as a graphical representation of data.
As such, they appear most frequently in our corpus.
It is common to author and store visualization as \textbf{raster graphics} (bitmaps) for easy usage and sharing~\cite{satyanarayan2019critical}.
Nevertheless, 
raster graphics are a standalone and lossy representation which loses the visualization semantics (\eg chart type, visual encoding, underlying data).
To perform automated analysis,
reverse engineering is often a pre-requisite, \ie to reconstruct the lost information using computer vision and machine learning approaches~\cite{poco2017reverse,savva2011revision}.
However, reverse engineering still remains an open problem with challenges to overcome in terms of robustness and accuracy~\cite{fu2020chartem}.
In conclusion, the lossy nature of raster graphics hinders the machines from easily interpreting and transforming the visualization~\cite{fu2020chartem}.

\textbf{Vector graphics} represents a less lossy alternative.
They have advantages over raster graphics in that they can be scalable up without aliasing.
Visualizations are usually stored in the Scalable Vector Graphics (SVG) format~\cite{satyanarayan2019critical},
which allows describing visual elements as shapes (\eg rectangles and text) with styles (\eg positions and fill-color).
Those low-level descriptions reduce the difficulties of reverse engineering,
\eg it is no longer necessary to apply computer vision techniques to detect objects such as texts~\cite{Moritz2017}.
Besides, this format enables support for interactivity and animation.
Nevertheless,
high-level visualization semantics such as visual encoding and underlying data is still lost, whose extraction requires considerable effort~\cite{harper2014deconstructing,wu2020mobilevisfixer}.


\subsubsection{Program}
Researchers have developed approaches for describing and storing visualizations as computer \textbf{programs}.
Programs retain necessary information to construct the visualization,
\eg
the underlying data.
The information is usually represented by languages,
which are classified into imperative and declarative programming.

\textbf{Imperative} visualization languages require users to specify step-by-step commands to create the visualization.
For example,
D3~\cite{bostock2011d3} is a JavaScript-based toolkit that assists programmers in specifying graphical marks (\eg bar, line) with visual attributes.
Programs by imperative languages are open-ended and thus allow the flexible creation of visualizations.
However, the open-ended characteristic renders it unstructured, with irregularities and ambiguities that hinder efficient machine-analysis from extracting the semantics (\eg visual encodings).
As such,
we observe few efforts in analyzing imperative programs.
In a slightly different vein,
Bolte and Bruckner~\cite{bolte2020vis} recently proposed Vis-a-Vis, a visual analytic approach for exploring visualization source codes.
The authors argued for more efforts on analyzing visualization programs.

\textbf{Declarative} languages ask programmers to directly describe the desired results, which is referred as specifications.
Specifications (\eg Vega-Lite~\cite{satyanarayan2016vega}) encapsulate step-by-step commands for visualization reconstruction into semantic components such as data encodings, axes, and legend properties.
This encapsulation is achieved by providing sensible defaults and introducing constraints with prescribed properties and structures.
As such, 
declarative programs tend to be less or equally expressive as imperative programs,
depending on their design.
Since specifications contain tags or markers to separate semantic elements and enforce hierarchies,
they are deemed semi-structured and thus more helpful for computer processing.
It is, therefore, common to generate or collect specifications to conduct data-driven research (\eg~VizML~\cite{hu2019vizml}).

\subsubsection{Hybrid}
Recent research proposes several \textbf{hybrid} content formats that incorporate the benefits of both graphics and programs.
Although such efforts remain limited,
we provide our embryonic classification here,
hoping to motivate future theories and models.

Two approaches aim to \textbf{embed programs into graphics}.
VisCode~\cite{zhang2020viscode} presents an embedding approach based on deep image steganography, that is, 
to conceal visualization specifications and meta information within the bitmap image.
Similarly, 
Chartem~\cite{fu2020chartem} encodes information (\eg the visualization specifications) in the background of a chart image. 
Both embedding techniques reduce the overhead to decode underlying visualization specifications, 
by showing that the encoded information can be extracted in an efficient and less error-prone manner while reducing the interference to human perception.

Loom~\cite{raji2020dataless} takes a different approach and seeks to \textbf{organize graphics by programs}.
Loom proposes to share interactive visualizations by filling the gap between two extremes, 
\ie sharing non-interactive formats such as images, 
and sharing the data, source codes, and software.
Specifically, 
it formulates interface visualizations as a standalone object built on an action tree.
Each intermediate node of the tree represents an interaction such as clicking,
and the leaf node stores the resulting visualization image.
Therefore, users can share and interact with the graphics as if they have access to the original source codes and software.

\subsubsection{Discussion and Open Questions.}
Despite the advantages of vector graphics, they appear less often (9) than raster graphics (44),
which might be due to the popularity of raster graphics in daily life.
However,
the lossy nature restricts the availability of machine interpretation and requires continued research on reverse engineering.

Arguably, programs are not friendly to most people except for programmers.
As such, programs tend to be less shared by laypeople online,
which hinders their collection and reuse.
For instance,
existing corpora mainly include graphics (\eg~\cite{battle2018beagle,borkin2013makes,lee2017viziometrics}) or tabular datasets (\eg~\cite{hu2019viznet}) instead of programs,
although programs are also found to be common in our corpus (42/98).
This suggests the need for more recognition of balancing the machine- and human-friendliness of visualization formats.

Visualization data often appears as byproducts in different mediums such as PDF documents~\cite{davila2020chart}, computational sheets, and notebooks~\cite{rule2018exploration}.
In order for AI to access them,
it needs to be able to recognize and extract visualization data,
which brings challenges for interpreting the medium's structure, developing conventional standards, and proposing object detection algorithms.
To that end,
recent research in parsing and analyzing visualization notebooks seems a promising direction~\cite{oppermann2020vizcommender}.




%% file: section/data/D2Feature.tex
\subsection{Representation}
\label{sec:data:representation}

\begin{table*}[!ht]
\centering
\caption{A summary of feature representation of visualizations for machine learning tasks.}
\vspace{-1em}
\label{table:feature}
\begin{tabular}{|c|P{5.5cm}|c|c|c|}
\hline
                         \cellcolor{tableBG}    & \cellcolor{tableBG}\textbf{Graphics} & \cellcolor{tableBG}\textbf{Program} & \cellcolor{tableBG}\textbf{Text} & \cellcolor{tableBG}\textbf{Underlying Data} \\ \hline
\cellcolor{tableBG}\textbf{Feature Engineering} &  \parbox[t]{5.5cm}{general image descriptors~\cite{savva2011revision,lee2017viziometrics,choudhury2016scalable,ray2015architecture} \\ element positions or regions~\cite{choi2019visualizing,savva2011revision,poco2017reverse,battle2018beagle,huang2007system} \\ element styles~\cite{battle2018beagle}}                 &  \parbox[t]{2.5cm} { parameters~\cite{wu2020mobilevisfixer,qian2020retrieve} \\
communicative signals~\cite{burns2012automatically} \\ design rules~\cite{moritz2018formalizing,lin2020dziban}} &   \parbox[t]{3cm}{statistical models~\cite{oppermann2020vizcommender,choudhury2016scalable,kim2018multimodal,luo2020steerable}}            &   \parbox[t]{2.5cm}{statistics~\cite{hu2019vizml,hu2019viznet,luo2018deepeye,key2012vizdeck,luo2018deepeyekeyword} \\ 
one-hot vector~\cite{xu2018chart}}             \\ \hline
\cellcolor{tableBG}\textbf{Feature Learning}    &  \parbox[t]{5.5cm}{convolutional neural network~\cite{bylinskii2017learning,lin2018vizbywiki,ma2020ladv,choi2019visualizing,poco2017reverse,siegel2016figureseer,fu2019visualization,dai2018chart,kim2018multimodal,chagas2018evaluation,tsutsui2017data,fu2019visualization,tang2016deepchart,haehn2018evaluating,chaudhry2020leaf,chen2019neural,chen2020figure,kafle2018dvqa,kahou2017figureqa,methani2020plotqa,reddy2019figurenet,singh2020stl} \\ autoencoder~\cite{zhang2020viscode,fu2019visualization}}
 &    autoencoder~\cite{zhao2020chartseer,obeid2020chart}             &  embedding models~\cite{xu2018chart,oppermann2020vizcommender}             &   autoencoder~\cite{dibia2019data2vis}          \\ \hline
\end{tabular}
\vspace{-1.5em}
\end{table*}

Now that we have considered different content formats of raw visualization data,
the next challenge is its representation.
Firstly,
raw visualization data in the format of images or programs might not explicitly represent the semantic information needed (\eg chart type).
Thus,
it is usually helpful to store and operate visualization data in the internal representation formats to facilitate the process. 
Secondly,
raw data need to be converted into feature representation to enable machine-learning techniques.
Thus,
we discuss feature engineering and feature learning approaches used to extract visualization features.



\begin{figure}[!b]
    \vspace{-1em}
    \setlength{\abovecaptionskip}{0pt}
	\centering
	\includegraphics[width=1\linewidth]{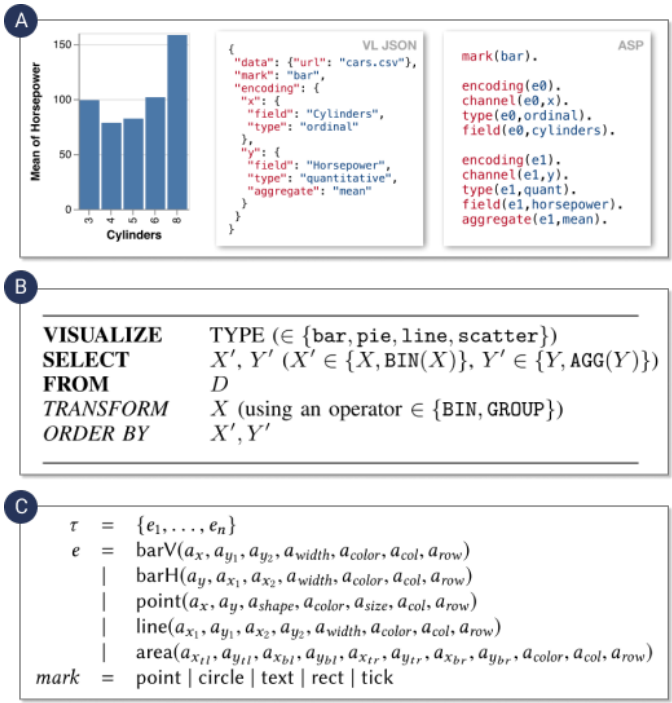}
	\caption{Internal representations of visualizations: {\protect\circled{A}} logic programming maps specifications into logic facts for computation~\cite{moritz2018formalizing}; {\protect\circled{B}} relational programming expresses visualizations in query language for relational operations such as selection~\cite{luo2018deepeye}; {\protect\circled{C}} set-theoretic programming facilitates set operations~\cite{wang2019visualization}.}
	\label{fig:partialSpec}
\end{figure}

\subsubsection{Internal Representation}
Visualization programs tend to contain extraneous details (\eg visual style) or miss semantics (\eg chart type) that might not meet particular needs of research.
Thus, systems usually express and operate on visualizations in simpler or more structured formats by removing unwanted or unnecessary specifications and adding customized information, which we refer to as \textbf{internal representations}.
However,
we only find 13 papers that explicitly propose formal internal representation formats that are designed towards the high-level goal of facilitating computation (\autoref{fig:partialSpec}). 


Draco~\cite{moritz2018formalizing} (\autoref{fig:partialSpec}\circled{A}) uses Answer Set Programming to express visualization specifications as logical facts.
This falls into \textbf{logic programming}~\cite{lifschitz1999action},
which expresses problem domains as facts or rules and benefits logical computation.
For instance, logical facts are used to check whether the specifications satisfy compound rules regarding the design knowledge.

Literature from data mining and databases~\cite{luo2018deepeye,luo2018deepeyekeyword,ehsan2017efficient,ehsan2016muve,mafrur2018dive,luo2020interactive,luo2020steerable,siddiqui2016effortless,vartak2015seedb,wu2017combining}
usually uses \textbf{relational programming} to form visualizations as queries into database (\autoref{fig:partialSpec}\circled{B}).
Those relational queries facilitate operations on collections of visualizations such as composing, filtering, and sorting.
This paradigm is also adopted by the CompassQL language in Voyager~\cite{wongsuphasawat2016towards,wongsuphasawat2015voyager,wongsuphasawat2017voyager}.

Finally,
Wang~\ea~\cite{wang2019visualization} proposed a \textbf{set-theoretic programming} representation that describes charts as a set of visual elements (\autoref{fig:partialSpec}{\protect\circled{C}}),
which facilitates set computation,
\eg to determine whether a visualization is a super-set of another.

\subsubsection{Feature Representation}
\label{sec:representation:features}
In this section, 
we discuss the features of visualizations,
which are the measurable properties serving as the input to machine learning models.
Features are extracted by \textbf{feature engineering} or \textbf{feature learning}~\cite{bengio2013representation}.
Feature engineering is the process of using domain knowledge to extract features from raw data,
while feature learning replaces this manual process by developing automated approaches that automatically discover useful representations.
The features of visualization data are multimodal,
including \textbf{graphics}, \textbf{program}, \textbf{text}, and \textbf{underlying data} (\autoref{table:feature}).
Some papers use multiple features for different tasks~\cite{choi2019visualizing,poco2017reverse} or use \textbf{hybrids} by feature fusion for improving performances~\cite{savva2011revision,choudhury2016scalable,kim2018multimodal,xu2018chart}.
In the following text, 
we describe each category in detail.

\textbf{Graphical} features are the most common features of visualizations.
The overarching goal is for predictive tasks, \eg to predict the chart type, or detection tasks.
Early work uses general image descriptors such as bag-of-keypoints~\cite{choudhury2016scalable} or patch descriptors~\cite{savva2011revision}.
These descriptors are designed for general visual content,
containing only low-level information such as shapes and regions.
To raise the level of abstraction,
researchers have proposed special domain descriptors for visualization-specific information,
and in many cases outperform general descriptors (\eg~\cite{poco2017reverse}).
Examples include regions of text elements (\eg titles and labels)~\cite{savva2011revision}, 
positions of text and mark elements~\cite{choi2019visualizing,poco2017reverse,battle2018beagle},
relative positions between text and marks~\cite{huang2007system},
and visual styles of elements~\cite{battle2018beagle}.
Despite promising results,
such feature engineering process remains labor-intensive and requires expertise.
Besides, 
it remains unclear whether human-crafted features are informative and discriminating for accomplishing machine learning tasks.
Therefore,
recent work has adopted feature learning more than feature engineering (\eg 12 versus 7 in 2020).
Particularly,
convolutional neural networks (CNNs) have been widely used to automatically learn spatial hierarchies of image features and shown to outperform early approaches~\cite{lin2018vizbywiki,ma2020ladv,choi2019visualizing,poco2017reverse,siegel2016figureseer,dai2018chart,kim2018multimodal,chagas2018evaluation,tsutsui2017data,tang2016deepchart,haehn2018evaluating,bylinskii2017learning,chaudhry2020leaf,chen2019neural,chen2020figure,kafle2018dvqa,kahou2017figureqa,methani2020plotqa,reddy2019figurenet,singh2020stl}.
VisCode~\cite{zhang2020viscode} recently uses an autoencoder to learn an effective representation that could conceal additional information.
Fu~\ea~\cite{fu2019visualization} used the latent vectors of the autoencoder model for prediction tasks.
However,
these models currently face the same challenge as early general image descriptors in that they might not capture visualization-specific information.
This limits their capability in high-level tasks such as assessment~\cite{fu2019visualization} and visual question answering~\cite{kafle2018dvqa}, where the performances are relatively dissatisfactory.

\textbf{Program} features are extracted from the programs such as specifications.
Probably the most straightforward representation is the parameters.
For instance,
MobileVisFixer~\cite{wu2020mobilevisfixer} and Retrieve-then-Adapt~\cite{qian2020retrieve} train models that learn to operate on the chart parameters, \eg positioning the element.
Burns~\ea~\cite{burns2012automatically} extracted communicative signals (such as whether a group of bars is colored differently from the other bars) that they fed into a Bayesian model.
Draco~\cite{moritz2018formalizing} contains constraints over facts that encode visualization design knowledge. 
These constraints describe whether a visualization conforms to best practices of effective visual design.
However, 
little work uses programs as the training input which might be due to several reasons, including the lack of training data or the overheads of reserve engineering to extract the program from visualization graphics.
Still, programs are a promising representation as they contain high-level visualization-specific information.
For example, ChartSeer~\cite{zhao2020chartseer} converts Vega-Lite specifications of charts into visualization embeddings by autoencoders.
The embeddings are used to measure similarities between charts to assist in analyzing chart ensembles,
and proven effective in user studies.
Their results suggest that program features are promising in semantically characterizing charts.

\textbf{Text} features refer to the text content in visualizations such as titles.
They are considered to improve the feature informativeness by incorporating semantic information.
For instance,
two systems describe text information with statistical models, \ie bag-of-words~\cite{choudhury2016scalable,kim2018multimodal}, and incorporate the resulting text features with graphical features to improve the performance for chart detection and classification tasks.
Moreover,
text features can capture the subject matter of visualizations.
VizCommender~\cite{oppermann2020vizcommender} is a recent content-based recommendation system built on machine learning models for predicting semantic similarity between two visualization repositories,
which shows a high agreement with a human majority vote.
Chart Constellations~\cite{xu2018chart} uses word embeddings to measure the similarity between charts.
However, it still remains unclear how text features could be effectively fused with graphical features to more comprehensively represent a visualization,
\eg to balance text-based and style-based similarities.

The last category of visualization features is in the \textbf{underlying data} that the visualization encodes. 
Chart Constellations~\cite{xu2018chart} describes the encoded data columns by a one-hot vector encoding,
which is then used to compute chart-wise similarities.
Besides such \textit{descriptive} purposes (\ie to describe a visualization),  
data features are found to be mainly \textit{predictive} (\ie to predict the visual encoding).
Data2Vis~\cite{dibia2019data2vis} adopts a sequence-to-sequence autoencoder structure that models the input dataset in the JSON format.
However, 
the sequence-to-sequence structure might not well capture the characteristics of the underlying data such as the data type.
DeepEye~\cite{luo2018deepeye,luo2018deepeyekeyword,luo2020steerable} and VizDeck~\cite{key2012vizdeck} perform feature engineering to consider data statistics such as the number of unique values in a column.
VizML~\cite{hu2019vizml} further extends this approach to 841 features including single- and pairwise-column features of the input dataset and significantly outperforms Data2Vis and DeepEye.
The analysis of VizML suggests that those features are not independent and some appear to be of little importance.
Future research should propose effective feature selection approaches or complementary feature learning methods.

\subsubsection{Discussion and Open Questions}
\textbf{Learning visualization-specific features}. Existing off-the-shelf computer vision or machine learning models are originally designed for general visual content or relational data. They might fall short when applied to visualizations, \eg visualization assessment~\cite{fu2019visualization} and Data2Vis~\cite{dibia2019data2vis} since they do not capture visualization-specific features such as visual encodings and encoded data that are critical. Researchers have proposed feature engineering approaches to craft features, which, however, are labor-intensive without guarantees for success. With the rapid development of deep learning techniques, we envision models tailored to visualizations that effectively learn visualization-specific features.

\textbf{Fusing multimodal features}. Visualizations are unlikely to be comprehensively represented by only one feature. However, as shown in~\autoref{fig:data}, most work only leverages graphic features while other modalities remain under-explored. In contrast, fusing multi-modal features could improve performances of chart detection and classification~\cite{choudhury2016scalable,kim2018multimodal}. Promising avenues for future work lie in leveraging feature fusion for high-level tasks such as question answering and similarity-based recommendation, which are not well-solved by the single-modal feature.




%% file: section/sec4Why.tex
\section{Goal: Why Apply AI to Visualization Data}
\label{sec:why}
\begin{figure}[!t]
    \setlength{\abovecaptionskip}{0pt}
	\centering
	\includegraphics[width=1\linewidth]{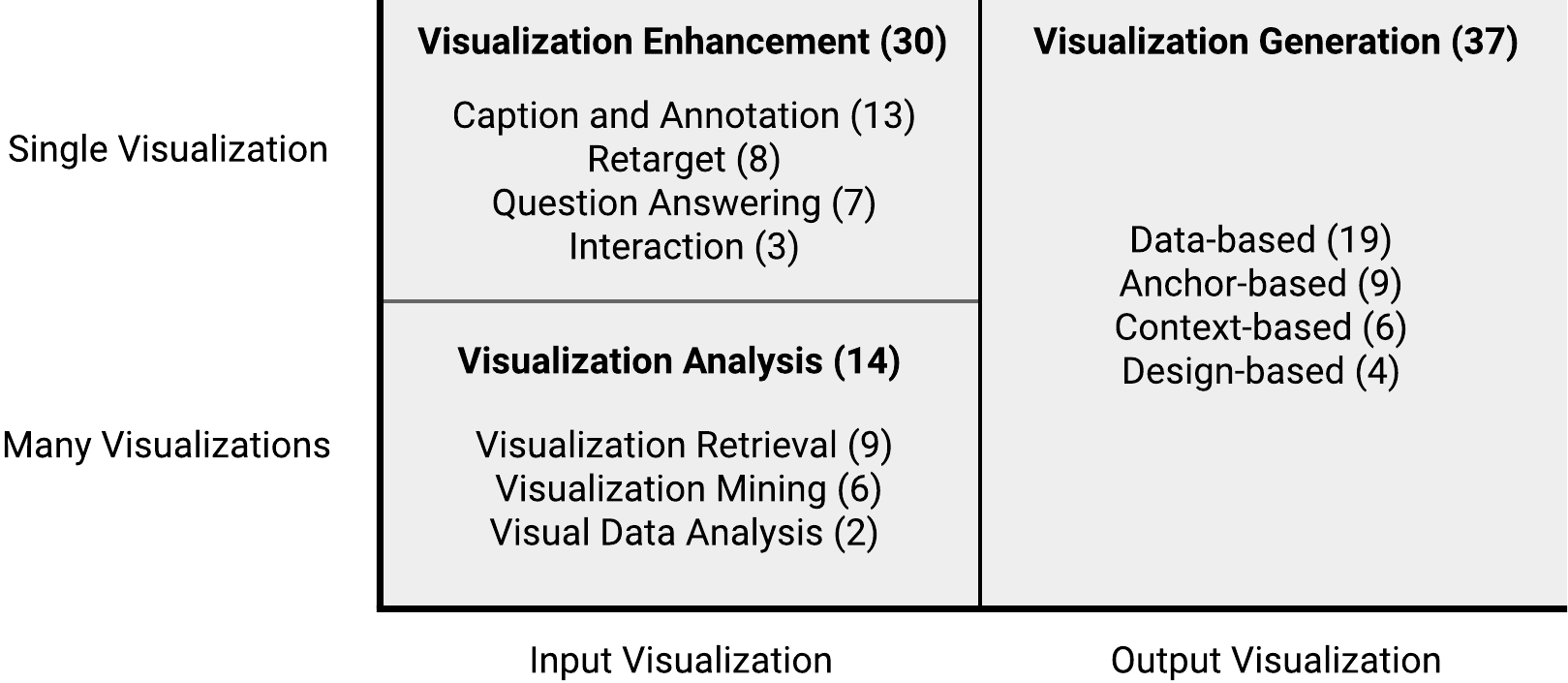}
	\caption{Matrix of goals of applying AI to visualization data with sub-categories. We subdivide goals along two axes deductively. Numbers in the parentheses indicate the count. A paper might have multiple goals.}
	\label{fig:goal}
	\vspace{-1em}
\end{figure}

The goals of applying AI to visualization data cover a wide spectrum,
pursued by research efforts from different areas.
We adopt a deductive classification method to create a mutually exclusive and collectively exhaustive taxonomy that better structures our discussion.
Specifically,
we subdivide goals along two axes deductively:
whether visualizations are the input or output and whether visualizations are single or many.
We further merge outputting single visualization and outputting many visualizations from an inductive perspective,
\ie
we observe that they share the same sub-categorization.
Therefore,
we finally classify goals into 3 categories, which are further subdivided (\autoref{fig:goal}):

\begin{compactitem}
    \item \textbf{Visualization Generation} outputs single or many visualizations given different user inputs.
    \item \textbf{Visualization Enhancement} processes and applies enhancement to an input visualization.
    \item \textbf{Visualization Analysis} concerns organizing and exploiting a visualization collection.
\end{compactitem}

It should be noted that some work is a technique paper that introduces a novel algorithm without practically demonstrating the applications (\eg~\cite{haehn2018evaluating}).
We do not label the goals for those techniques but instead discuss their tasks in~\autoref{sec:how}.


\input{section/goals/G1Generation}

\input{section/goals/G2Enhancement}
\input{section/goals/G3Database}



%% file: section/goals/G1Generation.tex
\subsection{Visualization Generation}
One of the central research problems in the visualization community is to ease the creation of visualizations,
which is mostly studied in surveyed papers (37 papers).
This is important since authoring effective visualizations is often time-consuming and challenging even for professionals~\cite{qin2020making}.
As such,
the ultimate goal of work in this category is to automatically generate visualizations.
We identified four subcategories that distinguish visualization generation approaches by user input.

\textbf{Data-based generation} outputs visualizations given a database or a data-table.
These approaches assist in visual data analysis and have been extensively studied over the last decades.
Early research dates back to 1986~\cite{mackinlay1986automating},
while it still remains an important question nowadays.
Recent work like Draco~\cite{moritz2018formalizing}, DeepEye~\cite{luo2018deepeye}, VizML~\cite{hu2019vizml}, and DataShot~\cite{wang2019datashot} make efforts on the direction.

Visual analysis is an iterative process where the next step of analysis often depends on earlier insights,
motivating the research on \textbf{anchor-based generation}.
The problem is to recommend a visualization given an anchor visualization.
For instance,
SeeDB~\cite{vartak2015seedb} intelligently recommends visualizations with large deviation to the anchor visualizations,
since they deem most ``interesting'' to users.
Similarly,
DiVE~\cite{mafrur2018dive} aims to vary the visualization recommendation,
while Dziban~\cite{lin2020dziban} targets at maintaining consistency.

Related to anchor-based generation,
\textbf{design-based generation} studies the problem of generating visualizations by injecting the target data into a reference design.
This is referred to as style transfer~\cite{harper2017converting} or visualization-by-example~\cite{wang2019visualization}.
Another recent example is Retrieve-Then-Adapt~\cite{qian2020retrieve} that applies pre-defined design templates to user information.

The last category is \textbf{context-based generation},
where the input only provides some contextual information such as a natural language description~\cite{cui2019text} or news articles~\cite{lin2018vizbywiki}.
An important task for context-based generation is to recommend data that is mostly related to the given context.

%% file: section/goals/G2Enhancement.tex
\subsection{Visualization Enhancement}
The proliferation of visualizations gives rise to research efforts in enhancing the use of existing visualizations (30 papers).
An important question is to \textbf{retarget} visualizations to different environments.
For example,
VisCode~\cite{zhang2020viscode} and Chartem~\cite{fu2020chartem} encode additional information in visualization images.
It is also common to summarize visualizations to generate natural language descriptions such as \textbf{captions} and \textbf{annotations}.
This approach transforms visualizations from visual to non-visual modality,
whereby enabling multimodal interactions~\cite{lai2020automatic} or enabling people with vision impairments to consume charts~\cite{choi2019visualizing}.
Related to natural language,
recent research challenges machines to perform \textbf{question answering} on visualizations,
that is, 
to generate answers given a question (\eg~\cite{kafle2018dvqa}).
Finally,
some work explores adding \textbf{interactions} to visualizations to improve the legibility and interactivity,
\eg Graphical Overlays~\cite{kong2012graphical} uses layered elements to aid chart reading.

%% file: section/goals/G3Database.tex
\subsection{Visualization Analysis}
With the increasing availability of visualization data,
recent research has constructed visualization databases and investigated methods for managing and analyzing these collections (14 papers).
\textbf{Retrieval} has been largely studied in the field of information systems and databases,
helping users search for visualizations that match their needs.
For instance,
Retrieve-Then-Adapt~\cite{qian2020retrieve} assists users in finding an example chart suitable for encoding their data.
Saleh~\ea~\cite{saleh2015learning} developed a search engine that returns stylistically similar visualizations given a query visualization.

Another promising set of work
has started to \textbf{mine} visualization collections to derive useful information
such as the visualization usage on the web~\cite{battle2018beagle} or in the scientific literature~\cite{ray2015architecture,lee2017viziometrics},
as well as design patterns in visualizations~\cite{hoque2019searching,smart2020color} and multiple-view systems~\cite{chen2020composition}.
The mined patterns provide evidence for recommending visualizations.

Related to mining,
some work facilitates \textbf{visual data analysis}.
Specifically,
they~\cite{dai2018chart,zhao2020chartseer} consider charts to be the analytical object and propose a visual analytic approach for analyzing data from charts ensembles by providing analytical guidance for effective visual analysis,
\eg which charts to examine next.

%% file: section/sec6Task.tex
\section{Tasks: How to Apply AI to Visualization Data}
\label{sec:how}
In this section, 
we focus on common tasks that researchers apply to visualization data.
We organize the observed tasks into seven primary tasks as follows where the number indicates the count:
\begin{compactitem}
    \item \textbf{Transformation} coverts visualization data from one modality (\eg graphics) to another (\eg program) (35/98).
    \item \textbf{Assessment} measures the absolute or relative quality of a visualization in terms of scores or rankings (19/98).
    \item \textbf{Comparison} estimates the similarity or other metrics between two visualizations (12/98).
    \item \textbf{Querying} finds the target visualization relevant with a user query within visualization collections (10/98). 
    \item \textbf{Reasoning} challenges AI to interpret visualizations to derive high-level information like insights and summaries (22/98).  
    \item \textbf{Recommendation} automates the creation of visualizations by suggesting data and/or visual encodings (38/98).
    \item \textbf{Mining} discovers insights from visualization databases (7/98).
\end{compactitem}

Most of those tasks originate from well-known terminology.
Transformation, assessment, and visual reasoning are well-studied tasks in the field of computer vision~\cite{computerVision},
while querying and mining come from database and information system research.
Two exceptions are comparison and recommendation.
Although the comparison is similar to the image similarity search task in computer vision~\cite{computerVision},
we find a large body of visualization research studying other metrics (\eg difference) between two visualizations and thus decide the wording comparison.
Recommendation is a widely studied task in the visualization literature (\eg~\cite{wongsuphasawat2016towards,zhu2020survey}).

We obverse relationships between tasks and published venues.
For instance, the Visualization community has mainly studied recommendation (17/34),
while Computer Vision and Artificial Intelligence research focuses more on transformation (8/16) and reasoning (7/16).
This shows the diversity and wide attention among different research communities.

\input{section/tasks/T0Transformation}
\input{section/tasks/T1Assessment}
\input{section/tasks/T2Compare}
\input{section/tasks/T3Query}
\input{section/tasks/T4Reasoning}

\input{section/tasks/T5Recommending}

\input{section/tasks/T6Mining}




%% file: section/tasks/T0Transformation.tex
\subsection{Transformation}
\label{sec:task:transformation}
Transformation is the operation that converts the content formats of visualizations.
Particularly,
it is straightforward to transform visualization programs into graphics by visualization tools or libraries (\eg \cite{bostock2011d3,satyanarayan2016vega}).
A more challenging problem is the reverse process, 
\ie reconstructing programs from graphics.
This process is also known as reverse engineering~\cite{poco2017reverse}.
In the following text,
we focus on the reverse engineering problem.

\textbf{Relations to goals and other tasks.}
Transformation is usually the first task for visualization enhancement and analysis,
especially when the input is images.
As such,
it is often a prerequisite for remaining tasks.
For instance,
the extracted information can be used for querying~\cite{hoque2019searching} and reasoning~\cite{kim2020answering}.

\textbf{Relations to visualization data.}
Several works study the problem of transforming visualization images into another image by altering the data or visual styles~\cite{qian2020retrieve,hoque2019searching,harper2017converting}.
Nevertheless,
their approaches are built on reverse engineering, 
\ie to extract the encodings first and then replace the data.
Little research explored the direct transformation in the image space~\cite{fu2020chartem,zhang2020viscode,brosz2013transmogrification,zhang2020dataquilt}.
As such,
we do not dedicate a separate discussion on image-to-image transformation at the current stage.
Future research could extend our taxonomy with further development in this field.


\subsubsection{Challenges and Methods}
Visualization reverse engineering has been widely and extensively studied over the past decades.
Early research could date back to 2001 when Zhou and Tan~\cite{zhou2001learning} proposed a learning-based paradigm for chart recognition.
Since then, much research has been devoted to extracting semantic information from visualization images such as chart types, visual encoding, and underlying data.
Ideally, reverse engineering is expected to yield cycle-consistency; that is, its output should be able to re-generate the original visualization.
Despite promising preliminary results,
several challenges remain to be overcome since much work makes simplifying assumptions on the expected input or output.
For instance, 
several approaches take vector graphics as input~\cite{hoque2019searching, harper2017converting, harper2014deconstructing,wang2018narvis,lu2017interaction+,huang2007extraction,wu2020mobilevisfixer,battle2018beagle,ray2015architecture}, assuming the type of each visual element is available as SVG.
Most work is limited to a predefined set of chart types,
while little work~\cite{lin2018vizbywiki,hoque2019searching, harper2017converting, harper2014deconstructing,wang2018narvis,wu2020mobilevisfixer,ma2020ladv,poco2017extracting} applies to more bespoke visualizations.
Besides, a large body of research focuses on extracting a particular portion of information,
such as chart classification~\cite{lin2018vizbywiki,ma2020ladv,lee2017viziometrics,kim2018multimodal,chagas2018evaluation,tsutsui2017data,tang2016deepchart},
object separation~\cite{browuer2008segregating,ray2016curve}, 
and object clustering~\cite{lu2017interaction+,wang2018narvis,wu2020mobilevisfixer}.

\begin{figure}[!t]
    \setlength{\abovecaptionskip}{0pt}
	\centering
	\includegraphics[width=1\linewidth]{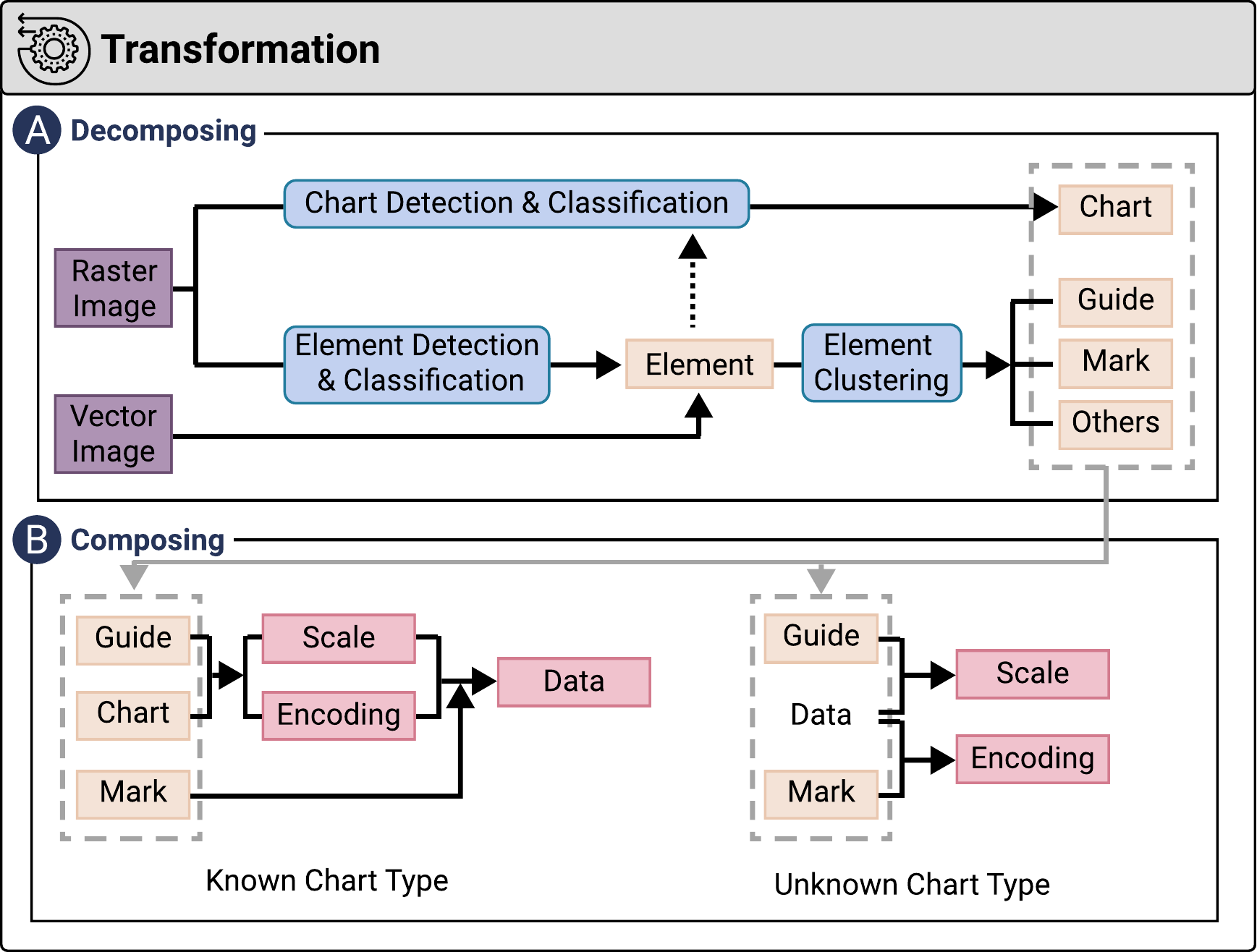}
	\caption{The conceptual framework of the reverse engineering process: {\protect\circled{A}} It first decomposes input graphics into semantic groups such as guides (axes and legends) and marks; {\protect\circled{B}} The resulting information is fed into mathematical computation to extract the visual encoding and/or underlying data, depending on whether the chart type is known. It remains an open challenge to derive both visual encoding and data from bespoke, unknown chart types.}
	\label{fig:reserseEng}
	\vspace{-1em}
\end{figure}

We summarize a conceptual framework of the reverse engineering process to provide an overview of existing approaches and identify research gaps.
We developed the framework via a bottom-up approach, 
where we abstract existing methods, identify their simplifying assumptions, and iteratively merge the results.
As shown in~\autoref{fig:reserseEng},
we identify two distinct phases.
The first phase decomposes visualization graphics into semantic elements (\eg axis, mark) through machine learning and computer vision techniques, including object detection, classification, and clustering.
The second phase performs mathematical computation over the decomposed semantic elements to extract visual encoding and/or the underlying data.
In the following text, we describe each phase.

\begin{table*}[!t]
\centering
\caption{The assessment task is classified by the output (row) and the method (column)}
\vspace{-1em}
\label{table:assessment}
\begin{tabular}{|c|c|P{4.1cm}|P{3cm}|}
\hline
\cellcolor{tableBG}        & \cellcolor{tableBG}\textbf{Rule-based} & \cellcolor{tableBG}\textbf{Machine Learning} & \cellcolor{tableBG}\textbf{Hybrid} \\ \hline
\cellcolor{tableBG}\textbf{Rankings} &  \parbox[t]{7cm}{ effectiveness rankings~\cite{mackinlay1986automating} }          &     \parbox[t]{4.1cm}{ learning-to-rank\cite{lin2018vizbywiki,luo2018deepeye,luo2018deepeyekeyword,luo2020steerable}}             &        \\  \hline
\cellcolor{tableBG}\textbf{Scores}   &  \parbox[t]{7cm}{convert rankings to scores~\cite{wongsuphasawat2015voyager,wongsuphasawat2017voyager,moritz2018formalizing}    \\ hand-crafted metrics~\cite{cui2019text,ehsan2016muve,ehsan2017efficient,bryan2016temporal,savvides2019significance,lee2019avoiding,zhang2020viscode,kim2020gemini,wu2020mobilevisfixer}
}&     \parbox[t]{4.1cm}{learning-to-rank with scores~\cite{qian2020retrieve} \\ predictive regression~\cite{key2012vizdeck,hu2019viznet,fu2019visualization} }  &    learning-to-weight hand-crafted metrics~\cite{moritz2018formalizing}    \\  \hline
\end{tabular}
\vspace{-1em}
\end{table*}

\textbf{Decomposing.}
The decomposing phase varies depending on the input (\autoref{fig:reserseEng}{\protect\circled{A}}).
The primary step for raster graphics is to detect and classify visual elements such as text and shapes.
This is approached by traditional image processing techniques (\eg edge detection, morphological operations) in early work~\cite{savva2011revision,choi2019visualizing,gao2012view,huang2007system} and machine learning or deep learning approaches (\eg Mark RCNN) in work published in 2015 and later~\cite{lai2020automatic,choi2019visualizing,chen2019towards,poco2017extracting, poco2017reverse,choudhury2016scalable,chen2015diagramflyer,siegel2016figureseer,dai2018chart,al2017machine}.
This element recognition step faces chart-specific challenges, \eg to cope with visual clutter in line charts and scatterplots~\cite{browuer2008segregating,ray2016curve}.
In addition, 
since existing object detection models are prone to rotation that is common for pie sectors,
Choi~\ea~\cite{choi2019visualizing} proposed a special heuristic to pie charts by grouping the nearby pixels with the same color.
The output of this element recognition step is usually the position and class of each visual element,
which are already available in the SVG specifications of vector graphics.
In other words, vector graphics remove the overhead of element recognition.

Chart detection and classification are an another step of the decomposing phase.
This step faces two important choices:
the classifier and the feature representation.
Classical classifiers (\eg support vector machine)~\cite{savva2011revision,battle2018beagle,choudhury2016scalable,prasad2007classifying} have been gradually superseded by deep learning classifiers (\eg convolutional neural network (CNN))~\cite{lin2018vizbywiki,ma2020ladv,choi2019visualizing,lee2017viziometrics,siegel2016figureseer,dai2018chart,kim2018multimodal,chagas2018evaluation,tsutsui2017data,tang2016deepchart} in visualization classification tasks.
This is mainly because CNNs can effectively learn abstract features from raw visualization images,
while classical classifiers require hand-crafted image features such as histograms of the image gradients (HOG)~\cite{prasad2007classifying} and dense sampling~\cite{savva2011revision,choudhury2016scalable}.
Several approaches seek to improve the representativeness of features by incorporating element-level features such as text~\cite{savva2011revision,kim2018multimodal} and shape style features~\cite{battle2018beagle}.
The last step of the decomposing phase addresses element clustering,
that is, to cluster visual elements into semantic groups, including guides (axes and legends), marks, and other information such as annotations.
This step is typically separately discussed for text and shape elements.
On one hand,
clustering text is usually formalized as a classification problem, that is,
to classify and group text according to their text roles such as x-axis-label and legend-title~\cite{al2016automatic,lai2020automatic,choi2019visualizing,savva2011revision,poco2017reverse,choudhury2016scalable,chen2015diagramflyer,dai2018chart,al2017machine,huang2007system}.
On the other hand,
this classification-based approach is not always readily applicable to shape clustering,
since the roles of shape depend on the chart type.
As such,
researchers usually simplify this problem by focusing on common charts where shapes are well-defined.
For instance,
Poco and Heer~\cite{poco2017reverse} trained a classifier to detect area, bar, line, and plotting shapes,
and consequently grouped shapes by types.
To support customized visualizations,
several approaches~\cite{wang2018narvis,wu2020mobilevisfixer,hoque2019searching, harper2017converting, harper2014deconstructing} use the node hierarchy information to group shape nodes under the same ancestor.
Nevertheless,
those approaches are only applicable to vector graphics.
Finally,
the shape clusters are associated with the text clusters to identify axes, legends, and label-mark relationships.

\textbf{Composing.}
After the visualization graphic has been decomposed into semantic groups (\eg guides and marks),
the \textbf{composing phase} (\autoref{fig:reserseEng}{\protect\circled{B}}) aims to extract the visual encoding and/or the underlying data from this semantic information.
Different from the decomposing phase that uses computer vision and machine learning tasks,
this composing phase mainly uses heuristics by leveraging domain knowledge about visualizations.
We find two common methodological themes of those heuristics,
depending on whether the chart type has been extracted from the previous phase.

The first class of heuristics uses information about the chart type and guides to determine the scale and the encoding,
which is dominant in our corpus~\cite{al2016automatic,lai2020automatic,choi2019visualizing,chen2019towards,savva2011revision,choudhury2016scalable,siegel2016figureseer,dai2018chart,al2017machine,gao2012view,huang2007system}.
For instance,
given a scatterplot with axes and legends,
it is straightforward to derive the visual encodings, \ie the x/y positions maps to numerical values and the color encodes the categorical data, and to calculate the scale.
Consequently,
the underlying data could be computed via applying the reserve scale computations over the marks.
However,
those heuristics are often limited to a small set of chart types.

Another class~\cite{hoque2019searching, harper2017converting, harper2014deconstructing} studies the more challenging problem of decoding bespoke visualizations where the chart type is unknown.
However,
they focus on D3 charts where the underlying data is available by crawling the SVG node on the web.
In this way,
they develop heuristics to determine the scale from guides and data, and to derive the encoding from data and marks.

\subsubsection{Discussion and Open Questions}
In conclusion,
it remains an open challenge to derive both the visual encoding and the underlying data from bespoke visualization graphics,
whose chart types are not limited to common ones.
An important future research direction would be to improve the current heuristics for determining visual encodings from bespoke visualizations,
\ie by machine-learning approaches.

Another primary challenge of reverse engineering lies in robustness and accuracy.
The pipeline of reverse engineering usually consists of multiple sequentially dependent tasks that are prone to single points of failure;
that is,
the failure of one task would spread to the whole system.
For instance, common failure cases such as text detection would impede the extraction of guides and consequently the visual encoding~\cite{poco2017reverse,lai2020automatic}.

This motivates the use of semi-automatic approaches that address imperfect algorithms with human intervention~\cite{kong2012graphical,mendez2016ivolver,jung2017chartsense}.
Nevertheless,
those frameworks of incorporating automatic algorithms with human intervention are specific to chart types and therefore not readily applicable to more general situations.
A related research direction is to investigate a general framework for deconstructing bespoke visualizations.

%% file: section/tasks/T1Assessment.tex
\subsection{Assessment}
There is a long history of research on teaching machines to assess and rank the quality of data visualizations.
As shown in~\autoref{table:assessment}, assessment outputs a numerical score of the visualization quality
or measures the relative quality in terms of ranking.

\begin{figure}[!h]
    \vspace{-0.5em}
    \setlength{\abovecaptionskip}{0pt}
	\centering
	\includegraphics[width=1\linewidth]{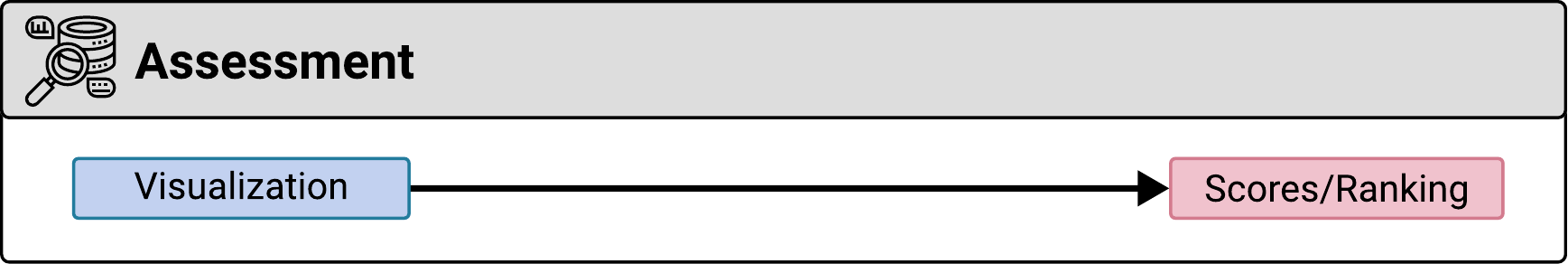}
	\caption{Input-output model of assessment.}
    \vspace{-0.5em}
\end{figure}

\textbf{Relations to goals and other tasks.}
The key motivation of assessment is to improve visualization design,
\eg to derive scoring metrics that can be used as cost functions for automatic generation.
That said,
assessment is often combined with recommendation. 

\textbf{Relations to visualization data.}
Most surveyed techniques take visualization programs as input,
focusing on the visual encoding and data quality.
Nevertheless,
Fu~\ea~\cite{fu2019visualization} propose an approach for assessing visualization images.

\subsubsection{Challenges and Methods}
Assessment is challenging due to the human-centered nature of visualizations that requires large-scale empirical experiments to understand what makes a visualization ``good''.
However,
the knowledge derived from large-scale empirical experiments is often represented as design guidelines instead of quantifiable rules.
As such,
much research aims to quantify knowledge about ``good'' visualizations.
In 1986,
Mackinlay~\cite{mackinlay1986automating} developed the APT system that \textbf{ranked} the effectiveness according to the accuracy rankings of quantitative perceptual tasks for different visual encoding channels.

However,
this ranking-based approach only reflects the relative quality of visualizations.
\textbf{Scoring-based} approaches are often more desirable since scores measure the absolute quality and therefore benefit down-streaming tasks, 
\eg scores can be used as the cost function for optimization.
To that end,
Voyager~\cite{wongsuphasawat2015voyager,wongsuphasawat2017voyager} and Draco~\cite{moritz2018formalizing} map different single-criteria rankings to numerical scores.
Besides,
researchers often leverage domain knowledge to design hand-crafted, \textbf{rule-based} metrics that measures the visualization quality such as informativeness~\cite{cui2019text}, interestingness~\cite{ehsan2016muve,ehsan2017efficient,bryan2016temporal}, accuracy~\cite{ehsan2016muve,ehsan2017efficient}, significance~\cite{savvides2019significance}, saliency~\cite{lee2019avoiding}, visual importance~\cite{zhang2020viscode}, complexity~\cite{kim2020gemini}, and mobile-friendliness~\cite{wu2020mobilevisfixer}.
However,
designing hand-crafted metrics usually requires considerable effort.
More critically, 
the design process is often unsystematic and lacks a strong methodological base.
For example, 
Wu~\ea~\cite{wu2020mobilevisfixer} demonstrated that even seemingly reasonable metrics do not always survive experimental scrutiny.
Besides, questions arise about how to weigh scores to reflect the overall, multi-criteria quality.
Several systems~\cite{cui2019text,wongsuphasawat2015voyager,wongsuphasawat2017voyager,kim2020gemini,ehsan2016muve,ehsan2017efficient} determine the weights for each score through manual refinements that could become unsystematic.

As such, 
another line of research seeks to propose more systematical \textbf{machine learning} approaches that learn to rank and/or score visualizations from data collected from empirical studies.
VizByWiki~\cite{lin2018vizbywiki} and DeepEye~\cite{luo2018deepeye,luo2018deepeyekeyword,luo2020steerable} formulate a learning-to-rank problem that learns to rank visualizations from crowdsourced data.
Retrieve-then-Adapt~\cite{qian2020retrieve} extends the learning-to-rank model that simultaneously outputs paired scores.
VizDeck~\cite{key2012vizdeck} learns a linear scoring function from users' up- and downvotes.
VizNet~\cite{hu2019viznet} demonstrates the feasibility of training a machine-learning model to predict the effectiveness of visual encodings.

Nevertheless, those machine-learning approaches face two major challenges, including poor generalisability and explainability.
First, 
the aforementioned models are trained over statistical or visual encoding properties,
assuming the underlying dataset and specification is available.
More importantly, 
they only support a limited number of chart types.
To that end,
Fu~\ea~\cite{fu2019visualization} propose a more general approach to assess the quality of visualization images that generalizes to different visualization type and does not require additional information except the visualization images.
Second,
machine-learning models lack explainability that might decrease trust.
Moreover, they often exclude knowledge derived from empirical studies.
To address these limitations, Draco~\cite{moritz2018formalizing} takes a \textbf{hybrid} perspective by encoding design knowledge as constraints and learning a weighting function to trade off those constraints, whereby outputting a final score.

\subsubsection{Discussion and Open Questions}
Existing approaches predominately focus on objective qualities that can be measured via user studies (\eg task performance and competition time).
However, 
subjective metrics such as aesthetics are relatively underexplored,
despite that they are considered as important features of good visualizations~\cite{McCandless09}.
This is challenging since it is difficult to harvest crowdsourced data of subjective quality of visualizations,
since crowdsourced judgments can be inconsistent and inaccurate.
This underscores research needs to propose methods for generating large-scale training datasets for visualization research in a reliable and sustainable manner.
One promising way is to incorporate expert knowledge and crowdsourcing experiments in dataset generation.

Going forward,
we envision machine-learning approaches that not only assess the visualization but also provide insightful explanations.
In this way,
the approaches embrace explainability by translating ML models into human-readable explanations and even useful design guidelines. 

%% file: section/tasks/T2Compare.tex
\subsection{Comparison}
Characterizing the similarity or other metrics between two visualizations is helpful when dealing with a visualization collection.

\begin{figure}[!h]
    \vspace{-0.5em}
    \setlength{\abovecaptionskip}{0pt}
	\centering
	\includegraphics[width=1\linewidth]{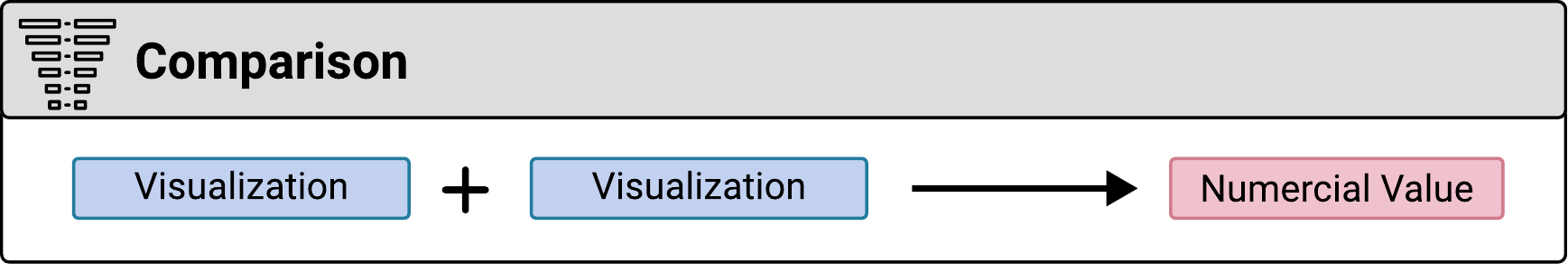}
	\caption{Input-output model of comparison.}
    \vspace{-0.5em}
\end{figure}

\textbf{Relations to goals and other tasks.}
Comparison is found to assist in visualization generation and analysis.
The comparison metrics can be used as cost-functions in 1) recommendation to perform anchor-based visualization generation (\eg~\cite{vartak2015seedb}),
as well as 2) querying to perform ``query-by-example'' (\eg~\cite{chen2020composition}).
Besides,
the assessment metrics can be used to compute the difference.

\textbf{Relations to visualization data.}
Comparison is studied on both visualization programs and graphics, as shown in~\autoref{table:comparison}.

\begin{table}[t]
\centering
\caption{Comparison is classified by the method (row) and the input (column)}
\vspace{-1em}
\label{table:comparison}
\begin{tabular}{|P{1.1cm}|P{1.2cm}|P{1cm}|c|P{2.3cm}|}
\hline
\cellcolor{tableBG}        & \cellcolor{tableBG}\textbf{Program} & \cellcolor{tableBG}\textbf{Text} & \cellcolor{tableBG}\textbf{Data} & \cellcolor{tableBG}\textbf{Graphics} \\ \hline
\cellcolor{tableBG}\textbf{Distance} & \cite{zhao2020chartseer,luo2020steerable,mafrur2018dive} & \cite{saleh2015learning,chen2020composition}          &     \cite{oppermann2020vizcommender,xu2018chart}             &  \cite{kandel2012profiler,law2018duet,luo2020interactive,vartak2015seedb,zhao2020chartseer,luo2020steerable,mafrur2018dive}  \\  \hline
\cellcolor{tableBG}\textbf{Difference}   & \cite{lin2020dziban,kim2017graphscape,xu2018chart} &      &    &  \\  \hline
\end{tabular}
\vspace{-1em}
\end{table}


\subsubsection{Challenges and Methods}
Perhaps the most straightforward approach for comparing two visualizations is to calculate the \textbf{difference}.
GraphScape~\cite{kim2017graphscape} is a directed graph model where each link represents an edit operation (\eg add field) and nodes denote the resulting visualizations.
Subsequently, 
each edit operation is registered with a cost,
which is learned from human judgments.
Dziban~\cite{lin2020dziban} further translates the graph model into a set of constraints and weights, similar to Draco~\cite{moritz2018formalizing}.
In this way,
both approaches explicitly model the difference between two charts as an operation associated with a numerical cost.
Nevertheless,
this difference-based approach becomes sophisticated when there exist multiple, often under-specified, operations between two charts,
where graph traversal is essential for searching and weighting all possible paths.
This is further complicated by the limitation that GraphScape only includes operations regarding data transformation and visual encodings.
Although it is methodologically feasible to extend GraphScape to support other operations such as recoloring,
such extensions are labor-intensive, without a guarantee for exhaustiveness.

Partly due to the above challenges of difference-based methods,
most research adopts \textbf{distance-based} measurements~\cite{chen2020composition,kandel2012profiler,law2018duet,luo2020interactive,luo2020steerable,mafrur2018dive,oppermann2020vizcommender,saleh2015learning,vartak2015seedb,xu2018chart,zhao2020chartseer}. 
The key idea is to convert a visualization into a feature vector, and compute the distance between two feature vectors according to distance functions.
Thus,
the technical challenges of distance-based measurements are two-fold: the choice of features and the distance function.

The features of visualizations vary among the underlying sources,
as discussed in \autoref{sec:representation:features}.
In the context of comparison,
we identify four primary sources including graphics, text, data, and specifications.
For instance,
Saleh~\ea~\cite{saleh2015learning} extract low-level visual features from \textbf{graphics} to learn style similarity,
and Chen~\ea~\cite{chen2020composition} model the configuration pattern of multiple-view visualization systems as a $1\times126$ vector measuring the layout.
Regarding \textbf{text} features, VizCommender~\cite{oppermann2020vizcommender} uses both hand-crafted features (\eg TF-IDF) and learned features (\eg Doc2Vec).
More work uses hand-crafted features for the \textbf{data}~\cite{kandel2012profiler,law2018duet,luo2020interactive,vartak2015seedb}.
Finally,
ChartSeer~\cite{zhao2020chartseer} uses deep learning approaches to convert Vega-Lite \textbf{specifications} into embeddings.

The derived features are subsequently fed into a distance function to derive the distance.
Examples of common distance functions include mutual information~\cite{chen2020composition,kandel2012profiler},
Earth Mover's Distance~\cite{luo2020interactive,vartak2015seedb},
Bhattacharyya coefficient~\cite{law2018duet},
and Jaccard coefficient~\cite{mafrur2018dive}.
Nevertheless,
the process of selecting distance functions is hardly detailed in the literature,
leaving rationales and insights unexplored.
This is worsened by the potential downside of distance-based measurements that the feature representation is less interpretable than the operations in difference-based measurements.
Thus,
it is often difficult to interpret the results,
and the user study suggests that the similarity measurement ``does not fully understand their (users') intent''~\cite{zhao2020chartseer}.

An important question that then arises is how to measure the overall similarity when combining multiple sources.
Naive feature concatenation is a natural way to combine features from different sources~\cite{mafrur2018dive,luo2020interactive}.
For instance,
Luo~\ea~\cite{luo2020steerable} propose a feature vector concatenated from five aspects, each represented by a one-hot vector describing visualization types, x-axis, y-axis, group/bin operations, and aggregation functions.
Another method is to compute the aggregated distance by weighting \textbf{hybrid} distances,
\eg chart encoding distance, keyword tagging distance, and dimensional interaction distance by Xu~\ea~\cite{xu2018chart}.
Nevertheless,
both feature concatenation and distance aggregation assume a linear relationship among different vectors,
which seems far from capturing the real-world complexity and thus yield limited performances when perceived by users.

\subsubsection{Discussion and Open Questions}
Comparison and assessment are closely related and share the same goal of outputting a numerical score. 
Comparison has been predominately rooted in feature engineering and hand-crafted distance functions.
The major downside is that it does not actually learn from user feedback and thus usually fails to meet the users' intent.
Unlike assessment,
few little machine learning approaches have been applied to comparison.
Several approaches (\eg~\cite{xu2018chart}) use pre-trained ML models to perform feature learning.
However,
they do not fine-tune the models on user feedback data.
Thus,
proposing dataset and ML approaches for comparison is a clear step to improve the performance.

Nevertheless,
it is non-trivial to adapt ML approaches to comparison,
since comparison involves two visualizations while standard ML models only take one entity as input.
ScatterNet~\cite{ma2018scatternet} addresses this issue. It is a deep learning model for predicting similarities between scatterplots by learning from crowdsourced human feedback data.
Nevertheless,
it is unclear how to adapt this approach to other statistical charts.
A key challenge is that CNN models used in ScatterNet are worse at capturing human perception in other charts~\cite{haehn2018evaluating,fu2019visualization}.


%% file: section/tasks/T3Query.tex
\subsection{Querying}
Querying is the task of retrieving relevant visualizations that satisfy the users' needs from a visualization collection.
It is a crucial component of Information Retrieval (IR) systems, which are also known as search engines, especially in the context of the web~\cite{cerulo2004taxonomy}.
Querying in this context is distinct from visualization query language (\autoref{fig:partialSpec}\circled{B}).
The latter specifies visualizations as a query into a database,
while the former describes a query into a visualization collection.

\begin{figure}[!h]
    \vspace{-0.5em}
    \setlength{\abovecaptionskip}{0pt}
	\centering
	\includegraphics[width=1\linewidth]{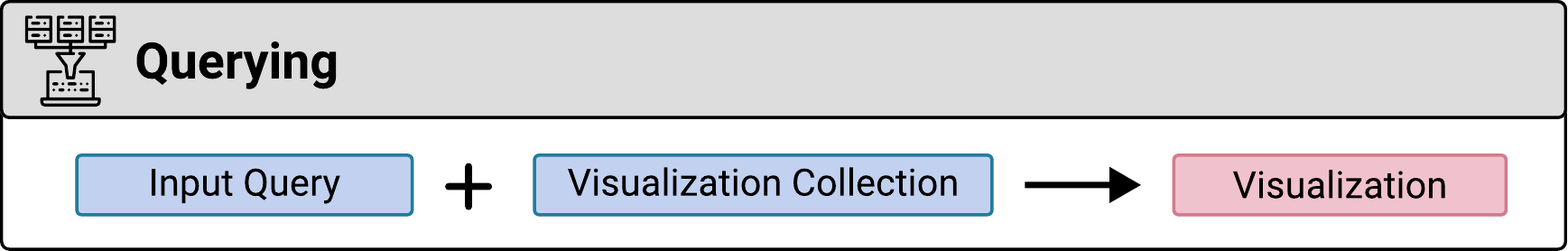}
	\caption{Input-output model of querying.}
    \vspace{-0.5em}
\end{figure}

\textbf{Relations to goals and other tasks.}
Querying is mainly for visualization retrieval (\autoref{table:querying}).
It is often built upon other tasks like transformation (\eg~\cite{hoque2019searching}) and comparison (\eg~\cite{chen2020composition}).

\textbf{Relations to visualization data.}
Querying directly in the image space can be difficult since semantic information is lost.
As such,
it is often performed on visualization programs,
where semantic information such as titles and axis labels are available.

\begin{table}[t]
\centering
\caption{Querying is classified by the method (row) and the input (column)}
\vspace{-1em}
\label{table:querying}
\begin{tabular}{|P{1.1cm}|P{1.5cm}|P{1.5cm}|c|c|}
\hline
\cellcolor{tableBG}        & \cellcolor{tableBG}\textbf{Keywords} & \cellcolor{tableBG}\textbf{Natural Language} & \cellcolor{tableBG}\textbf{Structural} & \cellcolor{tableBG}\textbf{Example} \\ \hline
\cellcolor{tableBG}\textbf{Exact} &  &          &   \cite{chen2020composition,hoque2019searching,siegel2016figureseer}           &    \\  \hline
\cellcolor{tableBG}\textbf{Best}   & 
\cite{ray2015architecture,srinivasan2018augmenting}
& \cite{li2014infographics,li2015novel}     & 
\cite{qian2020retrieve,chen2015diagramflyer} & 
\cite{saleh2015learning}
\\  \hline
\end{tabular}
\vspace{-1em}
\end{table}


\subsubsection{Challenges and Methods}
There are two viewpoints that characterize querying: how to specify users' needs and how to match the needs.

The simplest form of querying syntax is \textbf{keywords}.
Keywords are popular since they are intuitive and easy to express.
Choudhury and Giles~\cite{ray2015architecture} developed a search engine that allowed users to search figures by keywords in the captions.
Similarly, Voder~\cite{srinivasan2018augmenting} supports keyword-based queries into data fields as well as general words like `outlier' from the data fact associated with a visualization.
Li~\ea~\cite{li2014infographics,li2015novel} extends keywords to \textbf{natural language} queries by extracting structural keywords from queries and matching the extracted words with text in visualizations.
Nevertheless,
keywords-based queries often fail to disambiguate unstructured queries since words have multiple meanings.

\textbf{Structural} queries are a mechanism for resolving ambiguity and improving the retrieval quality
and have been used in multiple systems~\cite{chen2015diagramflyer,chen2020composition,hoque2019searching,qian2020retrieve,siegel2016figureseer}.
They are built on keywords with the addition of structural constraints.
For instance,
DiagramFlyer~\cite{chen2015diagramflyer} is a search engine where a query contains eight key structural fields (\eg type, x-label, legend) that can uniquely describe a visualization.
Those structural constraints make it possible to search information more than text in visualizations.
For instance,
Retrieve-then-Adapt~\cite{qian2020retrieve} retrieves infographics based on a query composed of graphical and textual elements.
Notably,
visualization specifications are a functional candidate for structural queries.
Hoque and Agrawala~\cite{hoque2019searching}'s search engine indexes D3 visualizations as Vega-Lite specifications
and supports queries in the Vega-Lite syntax. 
Accordingly,
it lets users find visualizations based on a wide range of constraints such as mark types, encodings, and non-data-encoding attributes.
In the best case where the input query is a complete Vega-Lite specification,
their research engine actually supports ``query-by-example''.

This \textbf{example-based} query is another format that offers an intuitive method for users to specify their intent.
Saleh~\ea~\cite{saleh2015learning} implemented a search engine for stylistic search over infographic corpora by returning stylistically similar images given a query image.
However,
more sophisticated analysis methods are necessary to capture characteristics beyond stylistic similarities.

Now that we have discussed the query syntax,
the next challenge is how to reason which visualizations are most relevant to the user-input query,
which can be classified into \textbf{exact-match} and \textbf{best-match} methods.
Exact-match techniques are used for filtering visualizations by strict conditions,
\eg to retrieve visual analytic systems containing four views~\cite{chen2020composition}, bar charts~\cite{hoque2019searching}, or charts describing a dataset~\cite{siegel2016figureseer}.
However, exact-matching is not always possible especially when the input conditions are too strict.
As such,
more systems use best-match approaches~\cite{chen2015diagramflyer,li2014infographics,li2015novel,qian2020retrieve,ray2015architecture,srinivasan2018augmenting,saleh2015learning} that rank visualizations according to metrics.
Those metrics measure the degree to which a visualization is relevant to the input query.
They use natural language models for text components  (\eg synonyms~\cite{chen2015diagramflyer,srinivasan2018augmenting}, text relevance~\cite{li2014infographics,li2015novel}) and similarity metrics~\cite{qian2020retrieve,chen2020composition}.
However,
relevance or similarity metrics are insufficient for retrieving visualizations that are not only relevant but also of high quality.
In response,
Retrieve-then-Adapt~\cite{qian2020retrieve} learns the distribution of visual elements in the corpus in an attempt to empathize visualizations with common elements,
assuming that the more frequent an element is, the better it is.

\subsubsection{Discussion and Open Questions}
Research on indexing visualizations has been relatively limited in the past decade,
leaving room to boost technical development.
Due to the diversity of visualizations,
even state-of-the-art methods are often restricted to certain types of visualizations,
\eg proportion-related infographics~\cite{qian2020retrieve} or basic D3 visualizations~\cite{hoque2019searching}.
How to generalize those approaches to more types of visualizations is a non-trivial issue that requires a deeper understanding of how visualizations should be indexed.
For instance, 
Hoque and Agrawala's approach~\cite{hoque2019searching} indexes D3 visualizations in Vega-Lite syntax,
which is insufficient in expressing user queries such as ``sunburst diagrams''.
Indexing becomes more challenging when applying to raster graphics (bitmap images) that deserve research efforts, \eg reverse engineering.

The intention gap is another challenge for querying visualizations.
From a theoretical perspective,
there are not enough empirical studies to understand the user needs in searching for visualizations.
Such studies are formative approaches to motivate the design space of the indexes of visualizations.
In a related venue,
Oppermann~\ea~\cite{oppermann2020vizcommender} recently found that information seeking is a core task when browsing visualization repositories, and that users were more interested in content rather than styles.
Similar studies are needed to survey users and inform the design of future search engines for visualization.
From a practical perspective,
it is crucial to design convenient query interfaces that assist users in specifying their intent.
For instance,
it is easy for users to specify the region of interest on an example visualization and interactively browse the results.
Hoque and Agrawala's query-by-example feature seems a promising start.
We also notice a large body of research that studies query-by-pattern or query-by-sketch in time-series visualizations, \eg ~\cite{wattenberg2001sketching,fan2020sketch}.
Future work could generalize those query paradigms to all visualization types.

%% file: section/tasks/T4Reasoning.tex
\subsection{Reasoning}
Reasoning challenges machines to ``read charts made for humans''~\cite{ono2018should}.
Reasoning requires interpreting visualizations to derive high-level information such as insights beyond extracting visual encoding and data via reverse engineering.
Reasoning is distinct from assessment since reasoning usually outputs semantic information (\eg insights, text summaries) rather than a numerical score.
As shown in~\autoref{table:reasoning},
we find three common classes depending on the targeted output of the reasoning process: visual perceptual learning, chart summarization, and visual question answering.
We first describe each class separately,
followed by an organized discussion of existing methods and research gaps.

\begin{figure}[!h]
    \vspace{-0.5em}
    \setlength{\abovecaptionskip}{0pt}
	\centering
	\includegraphics[width=1\linewidth]{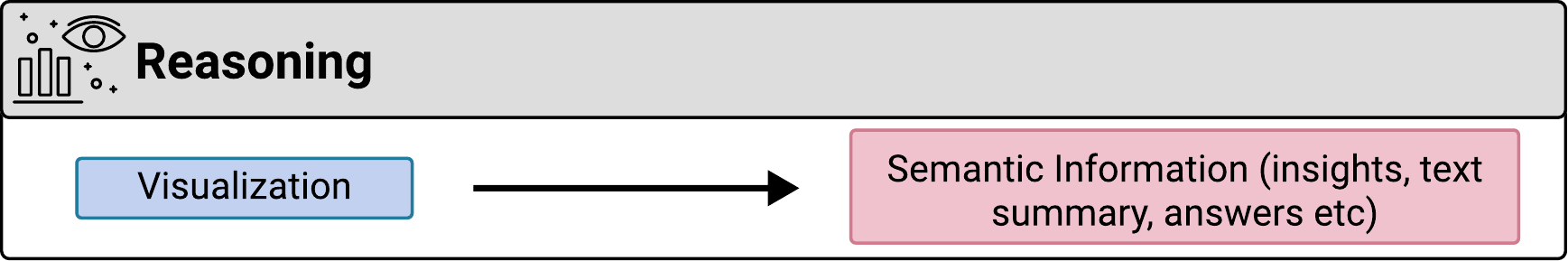}
	\caption{Input-output model of reasoning.}
    \vspace{-0.5em}
\end{figure}

\textbf{Relations to goals and other tasks.}
Reasoning is mainly for visualization enhancement (\eg summarize natural language descriptions~\cite{obeid2020chart}).
It sometimes relies on reverse engineering to improve the algorithm performance (\eg~\cite{kim2020answering}).

\textbf{Relations to visualization data.}
Reasoning is studied on both visualization images and programs.

\begin{table}[t]
\centering
\caption{Reasoning is classified by the class (row) and the method (column)}
\vspace{-1em}
\label{table:reasoning}
\begin{tabular}{|c|P{1.9cm}|P{2.35cm}|}
\hline
\cellcolor{tableBG}        & \cellcolor{tableBG}\textbf{Rule-based} & \cellcolor{tableBG}\textbf{Learning-based}  \\ \hline
\cellcolor{tableBG}\textbf{Visual Perceptual Learning} & \cite{bryan2016temporal}  & 
\cite{bylinskii2017learning,haehn2018evaluating}
\\  \hline
\cellcolor{tableBG}\textbf{Chart Summarization}   & 
\cite{mittal1998describing,luo2018deepeyekeyword,burns2012automatically,choudhury2016scalable,cui2019datasite,demir2012summarizing}
& \cite{liu2020autocaption,chen2019neural,chen2020figure,obeid2020chart}
\\  \hline
\cellcolor{tableBG}\textbf{Visual Question Answering}   & 
\cite{huang2007system,kim2020answering}
& \cite{chaudhry2020leaf,methani2020plotqa,kahou2017figureqa,reddy2019figurenet,kafle2018dvqa}
\\  \hline
\end{tabular}
\vspace{-1em}
\end{table}


\subsubsection{Challenges and Methods}
\textbf{Visual perceptual learning} aims to solve visual tasks by analyzing visual information.
For instance,
Temporal Summary Images~\cite{bryan2016temporal} automatically extracts points of interest in charts according to predefined heuristics.
Recently,
deep learning methods, trained on labeled datasets, have been applied to improve machine perception of visualization images.
Bylinskii~\ea~\cite{bylinskii2017learning} presented neural network models to predict human-perceived visual importance of visualization images.
Similarly,
Haehn~\ea~\cite{haehn2018evaluating} evaluated the performances of CNNs on Cleveland and McGill's 1984 perception experiments~\cite{cleveland1984graphical} and concluded that off-the-shelf CNNs were not currently a good model for human graphical perception.
Their initial results underscore the importance of continued research to improve the performance.

\textbf{Chart summarization} becomes increasingly important with the rapid popularization of visualizations.
Most existing approaches generate text summaries such as natural language description or captions~\cite{burns2012automatically,choudhury2016scalable,cui2019datasite,demir2012summarizing,luo2018deepeyekeyword,obeid2020chart,mittal1998describing,liu2020autocaption}.
The simplest approach is to provide a short description of how to interpret the chart~\cite{mittal1998describing,luo2018deepeyekeyword}, 
\eg ``This chart shows the trend of average departure delay in January''.
More advanced approaches focus on explaining and communicating high-level insights conveyed by charts.
Those approaches extract the data patterns and subsequently convert patterns to natural language according to pre-defined templates~\cite{burns2012automatically,choudhury2016scalable,cui2019datasite,demir2012summarizing},
\eg ``X was the most/least frequent sub-category in A''.
Liu~\ea~\cite{liu2020autocaption} offered an alternative perspective that learns most noteworthy insights with deep learning approaches.
A common limitation of the above work is that natural language summaries are generated via pre-defined templates and therefore confined to few variations and generality.
Thus,
recent research proposes several end-to-end deep-learning solutions for generating chart captions~\cite{chen2019neural,chen2020figure} or text summarization~\cite{obeid2020chart}.
However,
summarization still remains highly under-explored since the algorithm performances have much space for improvements.

\textbf{Visual question answering} is another emerging research area that aims to answer a natural language question given a visualization image.
Traditional methods~\cite{huang2007system} first decode visualizations into data tables and then parse template-based questions into queries over the data table to generate answers.
Kim~\ea~\cite{kim2020answering} recently improved a natural language parser to support free-from, crowdsourced questions.
Another line of research studies end-to-end deep learning approaches~\cite{chaudhry2020leaf,methani2020plotqa,kahou2017figureqa,reddy2019figurenet,kafle2018dvqa}.
The key challenge is that answering questions for visualizations requires high-level reasoning of which existing visual question answering models are not capable~\cite{kahou2017figureqa,kafle2018dvqa}.
Besides,
visualization images are sensitive to small local changes, 
\ie shuffling the color in legends greatly alters the charts' information.
Therefore,
the major problem is how to learn the features from visualization images and fuse them with features from natural language questions.
DVQA~\cite{kafle2018dvqa} learns and fuses features via a sophisticated model containing multiple sub-networks,
each responsible for different components such as spatial attention.
Later work expands this work with improvements to the models,
\eg PlotQA~\cite{methani2020plotqa} and LEAF-QA~\cite{chaudhry2020leaf} explicitly apply reverse engineering to retrieve visual elements and feed the extracted information into sub-networks.
However,
there still lacks real-world datasets, good models, and evaluation metrics.

\subsubsection{Discussion and Open Questions}
The reasoning task is currently undergoing changes since machine learning approaches, particularly deep learning models, are increasingly used.
This change may attribute to the rapid advancement and successful applications of deep learning in visual reasoning.
In the visualization context,
research gaps emerge since off-the-shelf models for natural images are often shown to yield dissatisfactory performances on visualizations (\eg ~\cite{haehn2018evaluating,kafle2018dvqa}).
This gap is not surprising since visualizations contain relational information that is sensitive to small details that are not commonly present in natural images,
\ie a local change to a bar shape might significantly impact the encoded data and conveyed meanings.
This research area remains largely under-explored.
First,
limited datasets are available, which hinders model development and validation.
For instance,
most existing datasets for visual question answering contain synthetic questions and charts,
which are far from being representative for the actual use.
Second,
it would be pertinent to study feature learning models that are tailored to visualization images.
The recent trend of decomposing end-to-end models to structures containing multiple sub-networks might be a promising method (\eg ~\cite{methani2020plotqa}). 

%% file: section/tasks/T5Recommending.tex
\subsection{Recommendation}
Recommendation is an important step for automating the creation of visualizations.
As shown in~\autoref{table:recommending},
there are three methods for recommending visualizations~\cite{wongsuphasawat2016towards}:
\begin{compactitem}
    \item Data recommendation suggests interesting data, insights, or data transformation be visualized from a database.
    \item Encoding recommendation determines the visual encoding (including both data and non-data encodings) given the data or other visualization elements.
    \item Hybrid recommendation decides both data and encodings.
\end{compactitem}

\begin{figure}[!h]
    \vspace{-0.5em}
    \setlength{\abovecaptionskip}{0pt}
	\centering
	\includegraphics[width=1\linewidth]{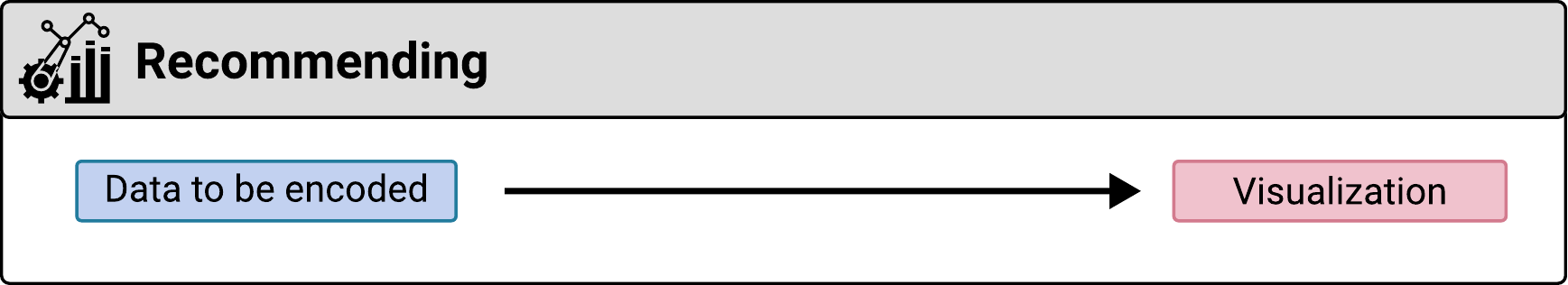}
	\caption{Input-output model of recommendation.}
    \vspace{-0.5em}
\end{figure}

\textbf{Relations to goals and other tasks.}
Recommendation is mainly for visualization generation.
It is related to assessment and comparison since the derived metrics are used as cost functions.

\textbf{Relations to visualization data.}
Recommendation outputs visualization programs and subsequently visualization graphics.


\subsubsection{Challenges and Methods}

\textbf{Data Recommendation.}
Given a dataset, one step of recommending visualizations is to select data fields to be visualized, and when applicable, corresponding data transformation as well.
The simplest approach to decide fields is enumeration.
For instance,
Voyager~\cite{wongsuphasawat2015voyager,wongsuphasawat2017voyager} enumerates all possible fields according to a predefined display order by the type and name.
DataSite~\cite{cui2019datasite} improves this enumeration approach by computing and communicating data facts associated with the selected fields according to pre-defined templates (\eg ``Correlation of A was found between X and Y'' if selected data fields are X and Y).
However,
enumeration imposes a heavy burden on users that motivates other research to recommend the most useful data facts, also known as insights.

\begin{table}[!t]
\centering
\caption{Recommendation is classified by the class (row) and methods (column)}
\vspace{-1em}
\label{table:recommending}
\begin{tabular}{|c|P{5cm}|P{1.2cm}|}
\hline
\cellcolor{tableBG}        & \cellcolor{tableBG}\textbf{Optimization} & \cellcolor{tableBG}\textbf{Prediction}  \\ \hline
\cellcolor{tableBG}\textbf{Data} & \cite{demiralp2017foresight,kandel2012profiler,srinivasan2018augmenting,wang2019datashot,luo2020interactive,shi2020calliope,ding2019quickinsights,mafrur2018dive,vartak2015seedb}  & 
\\  \hline
\cellcolor{tableBG}\textbf{Encoding}   & 
\cite{mackinlay1986automating,mackinlay2007show,wongsuphasawat2015voyager,ananthanarayanan2018datavizard,cui2019datasite,kandel2012profiler,narechania2020nl4dv,shi2020calliope,bouali2016vizassist,srinivasan2018augmenting,moritz2018formalizing,lin2020dziban,bryan2016temporal,cui2019text,kim2020gemini,ma2020ladv, wu2020mobilevisfixer,qian2020retrieve,smart2020color}
& \cite{dibia2019data2vis,hu2019vizml,wang2019datashot}
\\  \hline
\cellcolor{tableBG}\textbf{Hybrid}   & \cite{ananthanarayanan2018datavizard,cui2019datasite,kandel2012profiler,shi2020calliope,chen2020augmenting,wang2019datashot,wongsuphasawat2017voyager,wongsuphasawat2016towards,luo2018deepeye,luo2018deepeyekeyword,luo2020steerable}
& 
\\  \hline
\end{tabular}
\vspace{-1em}
\end{table}

Recommending insights is often approached by proposing a taxonomy of insight types (\eg extrema), each type associated with an assessment metric.
Examples are Foresight~\cite{demiralp2017foresight}, Profilier~\cite{kandel2012profiler}, 
Voder~\cite{srinivasan2018augmenting},
DataShot~\cite{wang2019datashot},
VisClean~\cite{luo2020interactive},
and Calliope~\cite{shi2020calliope}.
One key challenge that then arises is the assessment metric. 
Voder~\cite{srinivasan2018augmenting} introduces threshold-based heuristics that classify data facts into different tiers,
while the remaining systems propose various cost functions that better capture the differences between data facts.
Remarkably, 
QuickInsights~\cite{ding2019quickinsights} propose a unified formulation of insights and scoring metrics irrespective of the type.
In the context of anchor-based generation where the goal is to recommend data that meets some criteria with respect to anchor data,
the aforementioned assessment metrics are augmented by or replaced by comparison metrics in VisClean~\cite{luo2020interactive}, DiVE~\cite{mafrur2018dive}, and SeeDB~\cite{vartak2015seedb}.
For instance,
SeeDB~\cite{vartak2015seedb} recommends data by deviation with an anchor visualization.

Given the above metrics,
the next challenge is to compute the best or top-k insights.
This is challenging since the space of data facts grows exponentially with the number of data table columns.
As such,
researchers have proposed multiple strategies to speed up the computation.
For instance,
Foresight~\cite{demiralp2017foresight} uses sketching to quickly approximate the costs.
Other systems introduce efficient searching algorithms to recommend the top-k insights.
Those searching algorithms are primarily progressive or iterative~\cite{luo2020interactive,shi2020calliope,vartak2015seedb,mafrur2018dive,wang2019datashot}, outputting intermediate solutions that approximate the optimal one.
That said,
this data recommendation problem has not yet been formulated as a prediction problem that is solved by machine-learning models,
probably due to the lack of labeled training data.
Notably,
two approaches explicitly adopt tree-based algorithms~\cite{luo2020interactive,shi2020calliope},
leveraging the idea of GraphScape~\cite{kim2017graphscape} that the visualization design space can be modeled as a graph for greedy or dynamic programming.

Several systems recommend data according to other input beyond a database.
Particularly, those inputs are related to natural language that is beyond the core scope of this survey.
Examples include natural language statements in Text-to-Viz~\cite{cui2019text}, news articles in VizByWiki~\cite{lin2018vizbywiki}, keyword queries~\cite{luo2018deepeyekeyword}, 
and natural language interfaces (\eg NL4DV~\cite{narechania2020nl4dv} and FlowSense~\cite{yu2019flowsense}).

\textbf{Encoding Recommendation} decides data encodings and/or non-data encodings for styling (\eg positions).

Data encodings are extensively studied in the literature.
Early approaches date back to 1986 where the APT system~\cite{mackinlay1986automating} enumerates the visual encoding space and selects the ``best'' encodings according to the assessment ranking in terms of expressiveness and effectiveness.
This ranking-based recommendation is implemented and extended in later systems such as ShowMe~\cite{mackinlay2007show} and Voyager~\cite{wongsuphasawat2015voyager}.
In addition to ranking,
several systems propose heuristic rules to decide visual encodings given the insights extracted from data recommendation or visual tasks,
\eg extreme insights or finding extremes are mapped to histograms or scatterplots.
Examples include DataVizard~\cite{ananthanarayanan2018datavizard}, DataSite~\cite{cui2019datasite},
Profiler~\cite{kandel2012profiler},
NL4DV~\cite{narechania2020nl4dv},
Calliope~\cite{shi2020calliope},
VizAssist~\cite{bouali2016vizassist},
and Voder~\cite{srinivasan2018augmenting}.

The above heuristic-based data-encoding recommenders have recently been superseded by machine learning approaches due to the increasing availability of datasets.
Draco~\cite{moritz2018formalizing} and Dziban~\cite{lin2020dziban} learn to assess visualizations by weighting design rules for visualizations.
Subsequently,
they formulate a constraint optimization problem to recommend the best or top-k visualizations.
As discussed,
such approaches combining assessment and optimization are commonly adopted for recommending insights.
Other ML approaches seek to directly learn the mappings between data and visual encodings by training an end-to-end model,
including Data2Vis~\cite{dibia2019data2vis}, VizML~\cite{hu2019vizml},
and DataShot~\cite{wang2019datashot}.

In addition to data encodings, other approaches study how to recommend non-data-encoding attributes such as layouts and colors.
Most approaches formulate optimization problems with the primary goal to define the optimization target, that is, the assessment metrics. 
Several metrics are human-crafted cost functions~\cite{bryan2016temporal,cui2019text,kim2020gemini,ma2020ladv, wu2020mobilevisfixer},
while other metrics are data-driven,
including machine learning models trained on human assessment dataset~\cite{qian2020retrieve} and distances to common patterns mined from a corpus~\cite{smart2020color}.
Several systems contribute novel optimization algorithms to improve efficiency,
including reinforcement learning~\cite{wu2020mobilevisfixer} and Markov chain Monte Carlo methods~\cite{qian2020retrieve}.
Similar to recommending top-k insights, 
both algorithms are progressive.

\textbf{Hybrid Recommendation} decides both data and encodings.
A straightforward approach for hybrid recommendation is to combine data and encoding recommendation sequentially.
This approach is widely implemented in visualization recommenders including DataVizard~\cite{ananthanarayanan2018datavizard},
DataSite~\cite{cui2019datasite},
Profiler~\cite{kandel2012profiler},
Calliope~\cite{shi2020calliope},
Voder~\cite{chen2020augmenting},
and DataShot~\cite{wang2019datashot}.
Other approaches take an end-to-end perspective,
formulating the recommendation tasks as an optimization problem.
Examples are Voyager~\cite{wongsuphasawat2017voyager,wongsuphasawat2016towards}, DeepEye and its extensions~\cite{luo2018deepeye,luo2018deepeyekeyword,luo2020steerable}.
Thus,
the core problem is to provide an overall assessment score regarding both the data and encodings.
The first step is to translate a visualization into a formal representation, \ie visualization query language (VQL).
Assessment metrics are then proposed to evaluate the quality of a VQL representation.
Finally, 
Voyager ranks the recommended visualizations,
while DeepEye proposes efficient algorithms to generate the top-k visualizations.

\subsubsection{Discussion and Open Questions}
An ongoing discussion in recommending visualizations is the problem formulation including optimization and prediction.
Optimization requires the careful design of optimization functions which are often the assessment scores.
Another important concern is the efficient algorithms for solving the complex, sometimes multi-objective, optimization problem due to the huge design space of visualizations.
Generally speaking,
optimization-based approaches have the potential to be extended to the human-in-the-loop approaches,
since the predicted assessment scores help users determine the visualization quality.
Nevertheless,
hand-crafted cost functions for assessment are often insufficient.
On the other hand, machine learning assessment requires massive training data that labels human assessment and therefore is expensive.
In contrast,
prediction-based approaches such as VizML~\cite{hu2019vizml} only demand training data describing the dataset and visualizations without the need for human labeling.
This reduces the overhead of constructing dataset and thus helps boost the performance.
Thus, 
the above discussion opens up many interesting general questions: 
How to interpret the prediction-based ML models to understand the reasoning process and potentially derive assessment scores?
How to reduce the costs for collecting training datasets?

There exist other research gaps with respect to recommendation.
For instance,
few ML approaches have yet been applied to recommend insights and non-data-encodings.
Is it beneficial to collect training data for those tasks such as predicting important data fields given a data table?
Besides, 
many existing approaches only recommend top-k, separated visualizations given a large data table, while it is unlikely that ``top-k visualizations fit all''.
Future research should study how to recommend dashboards, and even visual analytics systems for more comprehensive and intelligent data analysis.
Generating a coherent data story compromising of multiple data facts is another interesting question.
Finally,
existing systems are confined to well-known charts.
Recent research in computer vision has proposed generative models for generating synthetic images.
An interesting question is to apply generative models to recommend synthetic, novel visualizations.

%% file: section/tasks/T6Mining.tex
\subsection{Mining}
Mining is an emerging task motivated by the rapid popularization and accumulation of visualization data online (\autoref{table:mining}).
Generally, there are two kinds of mining tasks, \ie mining design patterns and mining data patterns,
that are discussed in the following text.

\begin{figure}[!h]
    \vspace{-0.5em}
    \setlength{\abovecaptionskip}{0pt}
	\centering
	\includegraphics[width=1\linewidth]{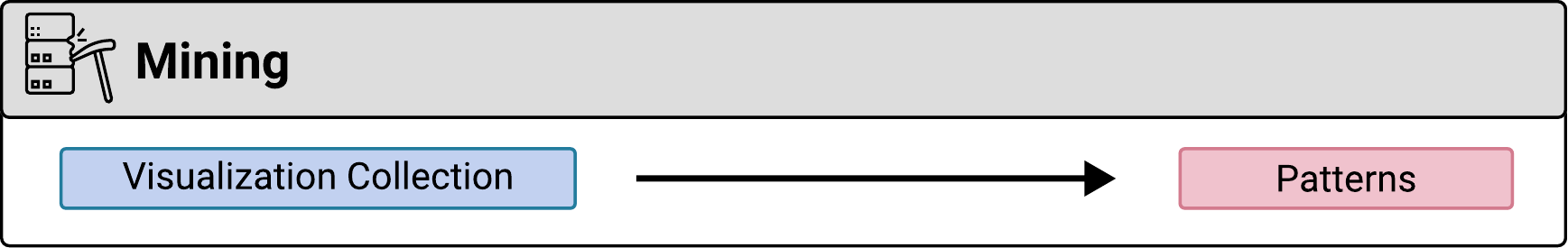}
	\caption{Input-output model of mining.}
	\vspace{-0.5em}
\end{figure}

\textbf{Relations to goals and other tasks.}
The goal of mining is visualization analysis, \ie to discover useful information or patterns from a visualization collection.
Reverse-engineering is often a prerequisite for mining to obtain semantic information.

\textbf{Relations to visualization data.}
Reasoning is studied on both visualization graphics and programs.

\begin{table}[b]
\vspace{-1em}
\centering
\caption{Mining is classified by the class (row) and the method (column)}
\vspace{-1em}
\label{table:mining}
\begin{tabular}{|c|P{2cm}|c|c|}
\hline
\cellcolor{tableBG}        & \cellcolor{tableBG}\textbf{Statistics} & \cellcolor{tableBG}\textbf{Clustering} & \cellcolor{tableBG}\textbf{Visual Analysis}  \\ \hline
\cellcolor{tableBG}\textbf{Design Pattern} & 
\cite{ray2015architecture,lee2017viziometrics,battle2018beagle,hoque2019searching,chen2020composition}
&  \cite{smart2020color} & 
\\  \hline
\cellcolor{tableBG}\textbf{Data Pattern}   & 
&  & \cite{xu2018chart,zhao2020chartseer}
\\  \hline
\end{tabular}
\end{table}

\input{tables/characteristics}


\subsubsection{Challenges and Methods}
\textbf{Mining Design Patterns.}
The concept of design mining refers to leveraging data mining techniques to derive design principles from existing artifacts.
Smart~\ea~\cite{smart2020color} formally introduced design mining in visualizations and proposed an unsupervised clustering technique to derive common color ramps.
However,
design mining has been implicitly practiced in several early systems but for different goals.
Choudhury and Giles~\cite{ray2015architecture} investigate the design patterns (\eg colored or not) of 300 line graphs sampled from top computer science conferences.
Viziometrics~\cite{lee2017viziometrics} computes the average number of visualizations in academic papers from different research domains.
In the visualization community,
Beagle~\cite{battle2018beagle} automatically crawled SVG-based visualizations online to investigate how popular,
\eg line and bar charts are on the web.
Hoque and Agrawala~\cite{hoque2019searching} performed a design demographics analysis of 7,860 D3 visualizations to identify common patterns such as how frequently circles are used.
More recently,
Chen~\ea~\cite{chen2020composition} studied the composition and configuration patterns in multiple-view visualizations collected from visualization papers.

The above systems mostly apply simple statistical analysis~\cite{ray2015architecture,lee2017viziometrics,battle2018beagle,hoque2019searching,chen2020composition} or clustering techniques~\cite{smart2020color}.
However, 
patterns might become meaningless numbers unless they are interpreted as insights or used to empower novel applications,
\eg to motivate design ideas.
As such,
several systems~\cite{hoque2019searching,chen2020composition} propose interactive visual interfaces for users to browse the visualization database and find examples.
Other approaches also leverage mined design patterns to recommend visual designs~\cite{smart2020color,chen2020composition}.

\textbf{Mining Data Patterns.}
Another line of research aims to explore the data patterns encoded in visualization ensembles.
This concept has been widely implemented in exploring visualization ensembles with respect to a specified type of charts (\eg ScagExplorer~\cite{dang2014scagexplorer} and TimeSeer~\cite{dang2012timeseer}).
We identify two approaches that are irrespective of the chart types,
namely Chart Constellations~\cite{xu2018chart} and ChartSeer~\cite{zhao2020chartseer}.
Both adopt a visual analytic approach by projecting charts into a 2D space whereby supporting clustering analysis and interactive analysis.

\subsubsection{Discussion and Open Questions}
There exist several promising directions for future work.
For design patterns,
existing approaches focus on visualization usage.
It would be beneficial to mine underlying semantic patterns from visualization collections,
\eg the relationships between linked views in a multiple-view visualization,
and the ``good'' or ``bad'' practices in existing visualization designs.
Those guidelines could motivate the design of recommender systems that automate the creation of visualizations.
For instance,
many recommender systems in our survey rely on manual coding to derive the design space of visualizations (\eg~\cite{wang2019datashot,wu2020mobilevisfixer,cui2019text}).
Automated mining of design space would significantly reduce the manual efforts.

Most current mining techniques are limited to simple statistics.
However, it is likely that there exist hidden patterns in visualization corpora.
Therefore,
one could explore the visualization collection with advanced mining techniques and human-in-the-loop analytics to uncover those patterns.
A research challenge would be to develop visual analytic systems for analyzing visualization collections.

%% file: tables/characteristics.tex
\begin{table*}[!ht]
\centering
\caption{A summary of characteristics of visualization data and research problems. Citations refer to example work instead of full instances.}
\vspace{-1em}
\label{table:characteristics}
\begin{tabular}{|p{0.45\linewidth}|p{0.49\linewidth}|}
\hline
\cellcolor{tableBG}\textbf{Characteristics of Visualization Data}   & \cellcolor{tableBG}\textbf{Research Problems} \\ \hline
It contains multimodal information such as programs, underlying data, and images. & 
How to represent visualization data~\cite{zhao2020chartseer}, translate from one modality to another~\cite{poco2017reverse}, and fuse and align information between modalities~\cite{chaudhry2020leaf}. \\ \hline
It is unnatural artifacts purposefully constructed with domain knowledge. &
How to automatically infer and/or incorporate with domain knowledge~\cite{moritz2018formalizing}. \\ \hline
It is susceptible to detailed local information such as marks' color and size. &
How to develop algorithms with higher precision and granularity~\cite{poco2017reverse} or involve human in the loop to fix errors made by algorithms~\cite{jung2017chartsense}. \\ \hline
It encodes data and is therefore data-dependent. &
How to handle data-dependent problems such as mathematical reasoning~\cite{kafle2018dvqa}.\\ \hline
\end{tabular}
\vspace{-1.5em}
\end{table*}

%% file: section/sec7FutureWork.tex
\section{Future Research Opportunities}
This survey describes the research vision of formalizing visualizations as a data format.
As shown in~\autoref{table:characteristics},
visualization data presents several differences from common data formats such as images and text.
Despite many research efforts,
there exist sufficient research gaps and potentials for future research.
In this section,
we outline an organized overview of future research directions.

\subsection{Visualization Standards and Interoperability}
Our analysis reveals different content formats of visualization data
that have been inconsistently adapted (\autoref{sec:data:rawdata}).
Such inconsistency impedes interoperability among different visualization systems. 
Being able to combine different systems and libraries is a common need in application settings.
For instance,
visualization generation tools such as VizML~\cite{hu2019vizml} recommend data encodings,
while other systems like MobileVixFixer~\cite{wu2020mobilevisfixer} adjust non-data encodings (visual styles).
Both systems do not work well together since their formats are incompatible with each other.
Notably,
it is a common practice to select partial information from the full specifications as the intermediate format.
This leads to the need for a common standard of visualizations that cover all existing partial formats,
as well as derivative tools for auto-completing partial specifications to generate universally compatible formats.

Besides,
systems for visualization enhancements usually have to engage in the reverse engineering to extract the underlying programs from graphics.
Despite extensive research efforts,
reverse engineering remains computationally expensive and lacks robustness~\cite{poco2017reverse}.
Particularly,
our analysis in \autoref{sec:task:transformation} suggests that it is currently impossible to perform reverse engineering on bespoke charts.
Although recent work like Chartem~\cite{fu2020chartem} and VisCode~\cite{zhang2020viscode} proposes new standards for storing programs in graphics,
there is a long way before such standards are adopted in existing systems.
Thus,
continued research on reverse engineering is essential and beneficial to interoperability.

\subsection{Visualization-Tailored Machine Learning}
Recent research has increasingly leveraged machine learning to generate or transform visualization.
However,
it remains challenging to choose the ``best'' representation and ML models for visualizations~\cite{wu2021learning}.
On the one hand,
programs are compact and effective representations that are computationally inexpensive~\cite{hu2019vizml,moritz2018formalizing}.
However,
programs might not generalize since they are often limited to specific chart types and might not apply to parameter values not observed during training.
On the other hand,
graphics appear to be a general representation.
However,
research suggests that off-the-shelf computer vision models for natural images achieve dissatisfactory results in tasks such as visual perceptual learning~\cite{haehn2018evaluating}, visual question answering~\cite{kafle2018dvqa}, and assessment~\cite{fu2019visualization}.

This gap is not surprising due to the characteristics of visualization images.
Compared with natural images, visualizations are sensitive to local details, 
\eg the sizes of graphical marks are of vital importance.
Text information is also critical, 
taking on various roles (\eg legends, axis labels) that have no tolerance for misinterpretation,
\eg missing a single character in axis labels leads to a great blunder.
Those characteristics make it challenging to design an ML model that are tailored to visualizations.
Recent work in visualization question answering (\eg~\cite{methani2020plotqa,chaudhry2020leaf,kafle2018dvqa}) has proposed sophisticated models to fuse multimodal features and capture visualization-specific features.
However,
such attempts are limited and their efficacy remains to be thoroughly evaluated.

Another important perspective lies in augmenting machine learning models with knowledge derived from empirical studies.
Machine learning models are often criticized for poor explainability.
Fortunately,
empirical research in the visualization field has accumulated a valuable knowledge base about how visualizations should be interpreted, assessed, and created.
It is therefore promising to incorporate that knowledge in ML models that is tailored to visualization research, \eg Draco~\cite{moritz2018formalizing}.

\subsection{A Big Data Perspective to Visualization Data}
This survey highlights the theme of considering visualizations as a data type.
Moving forward,
an evolving and promising theme is big visualization data,
which concerns processing and analyzing visualization data at larger scales.
This theme leads to several new issues that deserve research efforts.
Visualization database systems should effectively store and manage visualizations that are heterogeneous and often unstructured.
Moreover,
it is unclear how to mine and analyze big visualization data,
since existing approaches mainly focus on small-scale visualization collections and use simple statistics.
These challenges demand a holistic adaptation of data mining techniques to visualization data,
including but not limited to data cleaning, data transformation, data reduction, feature extraction and analysis.
For instance,
when collecting visualization datasets by web crawling,
it is important to clean the visualization collection by removing noise such as non-visualization images.
Therefore,
there are still many open research topics in terms of analyzing big visualization data.

\subsection{Human-Visualization Ecosystem}
\begin{revised}
Although this survey treats visualizations as data processed by AI,
we emphasize that visualizations are designed for and arguably central to humans.
Therefore,
it is important to establish the broader human-visualization ecosystem that AI could engage in and contribute to.
As discussed in~\autoref{sec:what},
there is a tradeoff between human- and machine-friendliness of visualization data formats that awaits better solutions.
It is also helpful to identify the limitations of AI approaches,
which offers possibilities of proposing mixed-initiative approaches (\eg~\cite{jung2017chartsense}).
Besides, it is promising to collect data regarding how humans perceive, assess and use visualizations at scale,
which empowers AI to learn new insights and enable learning-based methods for visualization and visual analytics~\cite{xu2020survey}.
Finally,
as many AI approaches focus on routine tasks (\eg querying),
a critical question arises of how AI could engage humans in creative practices such as creating novel visualizations (\eg~\cite{yuan2021infocolorizer}).\end{revised}


%% file: section/sec8Discussion.tex
\section{Discussion and Limitation}
\label{sec:discussion}
In this section,
we discuss the limitations in terms of mutual exclusiveness and collective exhaustiveness, as well as generalizability.

\subsection{Exclusiveness and Exhaustiveness}
In this survey,
we use an inductive approach to organize the literature and construct our taxonomy by observing existing work and iteratively generalizing the classification.

We note several dependencies among tasks.
For instance, assessment and comparison metrics are often used as optimization functions for recommendation.
However,
those dependencies should not be interpreted as violations of mutual exclusiveness.
Instead,
dependencies suggest that tasks can be sequentially combined into a \textit{system pipeline} for solving complex problems.

Because we inductively collected the papers for this survey,
we do not claim that our taxonomy is exhaustive.
There exist many potential research questions that we have not observed yet.
It would be interesting to improve the taxonomy from a deductive perspective by referring to task taxonomies in related fields such as computer vision, artificial intelligence, and databases.
For instance,
image compression and style transfer are well-studied tasks in the computer vision community~\cite{computerVision},
which, however, remain unexplored in the context of visualizations.
That said, 
there exist promising directions for future research of~\aivis{}.

\subsection{Generalization to Visualizations Beyond Charts}
This survey makes simplifying assumptions by focusing on charts and infographics,
excluding scientific visualizations and work tailored for a specific type of visualizations.
An important concern is how to generalize or extend the taxonomy to a wider spectrum of visualization data.
During our analysis process,
we find  
several exceptions that warrant future improvement.
In the following text,
we discuss notable extensions to our what-why-how taxonomy.

\textbf{What.} Our what taxonomy primarily focuses on visualization data.
However,
visualizations are hardly considered as standalone data in AI-empowered systems.
Instead,
it is often necessary to provide ground-truth labels for visualizations as auxiliary data, \eg chart types~\cite{poco2017reverse}.
Another type of auxiliary data is user-generated,
\ie interaction logs~\cite{fan2018fast} and analysis provenance~\cite{xu2020survey}.
Future research should study the taxonomy of auxiliary data to better contextualize the opportunities for AI for visualization research. 

\textbf{Why.}
The why taxonomy is organized along two axes: single or many visualizations versus inputting or outputting visualizations.
A missing perspective lies in work that neither inputs nor outputs visualizations but instead exploits visualization data in the middle stage.
For instance,
Lallé~\ea~\cite{lalle2015prediction} collected eye-tracking data when browsing visualizations and trained an ML model to predict the learning curves of visualizations.
Our survey does not cover this kind of research due to our theme centered on considering visualization as a data format,
emphasizing how visualization data are processed and produced.
Another perspective for improving our taxonomy is to expand the sub-categories by adding goals that are currently confined to a particular visualization.
For instance,
deep learning methods for brushing point visualizations~\cite{chen2019lassonet,fan2018fast} or drawing graphs~\cite{wang2019deepdrawing}
fall under visualization enhancement and generation category, respectively.

\textbf{How.}
Echoing the above discussion about ``what'',
there exist corresponding needs of identifying tasks for processing and analyzing auxiliary data.
This is crucial since our current task taxonomy concerns visualization data.
Moreover,
we notice another potential task for visualization data, 
namely \textit{visualization collection summarization} that aims to represent visualization collections in an effective and compact manner.
However,
existing approaches are limited and specific to visualization types,
\eg 
Scagnostics for scatterplots~\cite{dang2014scagexplorer} and line charts~\cite{dang2012timeseer}.
In addition to those statistical summarization approaches,
more recent visual analytic approaches use glyphs~\cite{zhao2020chartseer}.
Future work could provide a systematical overview of visualization summarization with increasing research efforts.

%% file: section/sec9Conclusion.tex
\section{Conclusion}
This survey probes the concept of considering visualizations as an emerging data format and investigate the advance of applying artificial intelligence to visualization data (AI4VIS).
We present a novel classification that enables the readers to find relevant literature among a wide variety of research areas.
Our classification can also help readers to understand current techniques and find areas for future research.
We hope that our survey could help stimulate new theories, problems, techniques, and applications.


%% file: section/biography.tex
\begin{IEEEbiographynophoto}{Aoyu Wu} is a Ph.D. student in the Department of Computer Science and Engineering at the Hong Kong University of Science and Technology (HKUST). He received his B.E. and M.E. degrees from HKUST. His research interests include data visualization and human-computer interaction.
For more details, please refer to \url{http://awuac.student.ust.hk/.}
\end{IEEEbiographynophoto}

\vspace{-10mm}

\begin{IEEEbiographynophoto}{Yun Wang}
received the BEng degree from Fudan University, and the PhD degree in computer science and engineering from the Hong Kong University of Science and Technology. She is a senior researcher at Microsoft Research. Her research interests are in data storytelling and visual data analytics. She has published papers extensively in IEEE VIS, the IEEE Transactions on Visualization and Computer Graphics, ACM CHI, etc. For more information, please visit \url{https://www.microsoft.com/en-us/research/people/wangyun/}.
\end{IEEEbiographynophoto}

\vspace{-10mm}

\begin{IEEEbiographynophoto}{Xinhuan Shu}
is currently a Ph.D. candidate in the Department of Computer Science and Engineering at the Hong Kong University of Science and Technology (HKUST). She received her B.E.degree  in  Computer  Science  and  Technology from Zhejiang University, China in 2017. Her research interests include data-driven storytelling, animated visualization, and visual analytics. For more details, please visit https://shuxinhuan.github.io/.
\end{IEEEbiographynophoto}

\vspace{-10mm}

\begin{IEEEbiographynophoto}{Dominik Moritz} is a professor at Carnegie Mellon University's Human-Computer Interaction Institute and a researcher at Apple. He has a PhD in Computer Science from the University of Washington. At CMU, he co-leads the Data Interaction Group. Dominik enhances people's ability to understand and communicate large and complex data with methods that integrate the capabilities of both people and machines. His website is at \url{https://www.domoritz.de/}.
\end{IEEEbiographynophoto}

\vspace{-10mm}

\begin{IEEEbiographynophoto}{Weiwei Cui}
received the BS degree in computer science and technology from Tsinghua University, China, and the PhD degree in computer science and engineering from the Hong Kong University of Science and Technology, Hong Kong. He is a principal researcher at Microsoft Research Asia, China. His primary research interest is visualization, with the focuses on democratizing visualization and AI-assisted design. For more information, please visit \url{https://www.microsoft.com/en-us/research/people/weiweicu/}.
\end{IEEEbiographynophoto}

\vspace{-10mm}

\begin{IEEEbiographynophoto}{Haidong Zhang}
received the PhD degree in Computer Science from Peking University, China. He is a Principal Architect at Microsoft Research Asia. His research interests include visualization and human-computer interaction.
\end{IEEEbiographynophoto}

\vspace{-10mm}

\begin{IEEEbiographynophoto}{Dongmei Zhang}
received the BE degree and ME degree from Tsinghua University, and PhD degree in Robotics from the School of Computer Science at Carnegie Mellon University. She is a Distinguished Scientist and Assistant Managing Director at Microsoft Research Asia, leading research in the area of Data, Knowledge, and Intelligence with research directions in data intelligence, knowledge computing, information visualization, and software engineering. For more information, please visit \url{https://www.microsoft.com/en-us/research/people/dongmeiz/}.
\end{IEEEbiographynophoto}

\vspace{-10mm}

\begin{IEEEbiographynophoto}{Huamin Qu} 
is a professor in the Department of Computer Science and Engineering (CSE) at the Hong Kong University of Science and Technology (HKUST) and also the director of the interdisciplinary program office (IPO) of HKUST. He obtained a BS in Mathematics from Xi'an Jiaotong University, China, an MS and a PhD in Computer Science from the Stony Brook University. His main research interests are in visualization and human-computer interaction, with focuses on urban informatics, social network analysis, E-learning, text visualization, and explainable artificial intelligence (XAI).
\end{IEEEbiographynophoto}